\documentclass[english,aps,pra,twocolumn,groupedaddress,superscriptaddress,nofootinbib]{revtex4-1}
\usepackage[T1]{fontenc}
\usepackage[latin9]{inputenc}
\usepackage{amsmath}
\usepackage{amssymb,amsbsy}
\usepackage{amsfonts}
\usepackage{graphicx}
\usepackage{hyperref}
\usepackage{color}
\usepackage{mathrsfs}
\usepackage{multirow}
\usepackage{float}
\usepackage{bm}
\usepackage{times,txfonts}

\makeatletter

\definecolor{nblue}{rgb}{0.3,0.3,1.0}
\definecolor{ngreen}{rgb}{0.2,0.7,0.2}
\definecolor{nred}{rgb}{0.9,0.1,0}
\definecolor{nblack}{rgb}{0,0,0}

\newcommand{\bra}[1]{\langle{#1}|}
\newcommand{\ket}[1]{|{#1}\rangle}

\newcommand{\beq}{\begin{equation}}
\newcommand{\eeq}{\end{equation}}
\newcommand{\bqa}{\begin{eqnarray}}
\newcommand{\eqa}{\end{eqnarray}}

\newcommand{\T}{^{\sf T}}

\newcommand{\gr}[1]{\boldsymbol{#1}}

\newcommand{\sig}{{\gr\sigma}}

\newcommand{\V}{\boldsymbol{V}}

\makeatletter

\usepackage{babel}
\begin{document}

\title{Investigating Einstein-Podolsky-Rosen steering of continuous variable \\ bipartite states by non-Gaussian pseudospin measurements}
\date{\today}

\author{Yu Xiang}
\affiliation{State Key Laboratory of Mesoscopic Physics, School of Physics, Peking University,
Collaborative Innovation Center of Quantum Matter, Beijing 100871, China}
\affiliation{Collaborative Innovation Center of Extreme Optics, Shanxi University, Taiyuan, Shanxi 030006, China}
\affiliation{Centre for the Mathematics and Theoretical Physics of Quantum Non-Equilibrium Systems (CQNE),
School of Mathematical Sciences,  University of Nottingham, Nottingham NG7 2RD, United Kingdom}

\author{Buqing Xu}
\affiliation{Centre for the Mathematics and Theoretical Physics of Quantum Non-Equilibrium Systems (CQNE),
School of Mathematical Sciences,  University of Nottingham, Nottingham NG7 2RD, United Kingdom}

\author{Ladislav Mi\v{s}ta, Jr.}
\affiliation{Department of Optics, Palack\'y University, 17. listopadu 12, 771 46 Olomouc, Czech Republic}

\author{Tommaso Tufarelli}
\affiliation{Centre for the Mathematics and Theoretical Physics of Quantum Non-Equilibrium Systems (CQNE),
School of Mathematical Sciences,  University of Nottingham, Nottingham NG7 2RD, United Kingdom}

\author{Qiongyi He}
\email{qiongyihe@pku.edu.cn}
\affiliation{State Key Laboratory of Mesoscopic Physics, School of Physics, Peking University,
Collaborative Innovation Center of Quantum Matter, Beijing 100871, China}
\affiliation{Collaborative Innovation Center of Extreme Optics, Shanxi University, Taiyuan, Shanxi 030006, China}

\author{Gerardo Adesso}
\email{gerardo.adesso@nottingham.ac.uk}
\affiliation{Centre for the Mathematics and Theoretical Physics of Quantum Non-Equilibrium Systems (CQNE),
School of Mathematical Sciences,  University of Nottingham, Nottingham NG7 2RD, United Kingdom}

\begin{abstract}
Einstein-Podolsky-Rosen (EPR) steering is an asymmetric form of correlations which is intermediate between quantum entanglement and Bell nonlocality, and can be exploited as a resource for quantum communication with one untrusted party. In particular, steering of continuous variable Gaussian states has been extensively studied theoretically and experimentally, as a fundamental manifestation of the EPR paradox. While most of these studies focused on quadrature measurements for steering detection, two recent works  revealed that there exist Gaussian states which are only steerable by suitable non-Gaussian measurements. In this paper we perform a systematic investigation of EPR steering of bipartite Gaussian states by pseudospin measurements, complementing and extending previous findings. We first derive the density matrix elements of two-mode squeezed thermal Gaussian states in the Fock basis, which may be of independent interest. We then use such a representation to investigate steering of these states as detected by a simple nonlinear criterion, based on second moments of the correlation matrix constructed from pseudospin operators. This analysis reveals previously unexplored regimes where non-Gaussian measurements are shown to be more effective than Gaussian ones to witness steering of Gaussian states in the presence of local noise. We further consider an alternative set of pseudospin observables, whose expectation value can be expressed more compactly in terms of Wigner functions for all two-mode Gaussian states. However, according to the adopted criterion, these observables are found to be always less sensitive than conventional Gaussian observables for steering detection. Finally, we investigate continuous variable Werner states, which are non-Gaussian mixtures of Gaussian states, and find that pseudospin measurements are always more effective than Gaussian ones to reveal their steerability. Our results provide useful insights on the role of non-Gaussian measurements in characterizing quantum correlations of Gaussian and non-Gaussian states of continuous variable quantum systems.
\end{abstract}
\date{\today}

\maketitle

\section{Introduction}\label{sec:intro}

Quantum information science is experiencing an intensive theoretical development and an impressive experimental progress, leading to revolutionary applications in computation, communication, simulation and sensing technologies \cite{Dowling}. Specific ingredients of quantum systems, including superposition phenomena and different manifestations of nonclassical correlations, are being harnessed for these tasks \cite{streltsov2016quantum,ent,disc,ABC,steeringreview1,steeringreview2,nonloc}. Characterizing the nature and degree of nonclassical correlations in quantum systems amenable to experimental implementation is thus of particular importance,  to assess their potential relevance as resources for quantum enhanced tasks.

Einstein-Podolsky-Rosen (EPR) {\it steering} \cite{schr,wiseman,steeringreview2} is a type of nonclassical correlations which is strictly intermediate between quantum entanglement \cite{ent} and Bell nonlocality \cite{nonloc}. Unlike the latter two, steering is asymmetric, meaning that a bipartite quantum state distributed between two observers Alice and Bob may be steerable from Alice to Bob but not the other way around. Originally regarded as a somehow puzzling manifestation of the EPR paradox \cite{epr,reid,Cavalcanti_09,steeringreview1}, steering is now appreciated as a resource \cite{resource} for a variety of quantum information protocols, including one-sided device-independent quantum key distribution \cite{branciard,walk,kogiasqss}, sub-channel discrimination \cite{PianiSt}, and secure teleportation \cite{Reid13,He15}.

Landmark demonstrations of EPR steering have been accomplished in particular in continuous variable (CV) systems \cite{handchen,seiji2015,walk,Deng17}, where nonclassical correlations can arise between degrees of freedom with a continuous spectrum, such as quadratures of light modes \cite{Brareview,serafozzi}. These experiments, as well as the majority of theoretical studies \cite{wiseman,wisepra,reid,steeringreview1,He2015,kogias,kogias2,kogiasqss,kogiasmono,Lami2016}, have focused specifically on verification and quantification of steering in so-called {\it Gaussian} states of CV systems as revealed by Gaussian measurements. This is motivated on one hand by the fact that Gaussian states, which are thermal equilibrium states of quadratic Hamiltonians, admit a simple and elegant mathematical description \cite{ourreview,pirandolareview,ournewreview}, and on the other hand by the fact that Gaussian states can be reliably produced and controlled in a variety of experimental platforms while Gaussian measurements are equally accessible in laboratory by means of homodyne detections \cite{serafozzi}.

However, it is necessary to go beyond the `small' world of Gaussian states and measurements in order to unlock the full potential of CV quantum information processing (e.g.~for universal quantum computation \cite{CVQC}), and to reach a more faithful characterization of the fundamental border between classical and quantum world. In this respect, two recent papers showed that there exist bipartite Gaussian states which are not steerable by Gaussian measurements, yet whose EPR steering can be revealed by suitable non-Gaussian measurements in certain parameter regimes \cite{OneWayPryde,NhaSciRep}. This opens an interesting window on the `big' non-Gaussian world and suggests that large amounts of useful nonclassical correlations could be overlooked by restricting to an all-Gaussian setting.

In this paper, we investigate EPR steering of two-mode CV states as detected by non-Gaussian measurements, specifically {\it pseudospin} measurements which have proven useful for studies of Bell nonlocality \cite{chen,chen2,gour1,gour2}. After setting up notation and basic concepts in Sec.~\ref{sec:prelim}, we begin our analysis by specializing to Gaussian states.  We consider in particular the prominent family of two-mode squeezed thermal states and identify regions in which their steerability can be detected by pseudospin measurements \cite{chen} (but not by Gaussian ones), using a steering criterion derived from the moment matrix \cite{kogias3} associated with such measurements. To accomplish this analysis, which goes significantly beyond the instances considered in the existing literature \cite{OneWayPryde,NhaSciRep}, we derive in Sec.~\ref{sec:fock} an explicit expression for the number basis representation of any two-mode squeezed thermal state, a result of interest in its own right. We further discuss an extension of our study to arbitrary two-mode Gaussian states by considering an alternative set of pseudospin operators \cite{gour1}, which are nevertheless found less effective than Gaussian measurements for steering detection. Our analysis of EPR steering in Gaussian states by either type of pseudospin measurements is presented collectively in Sec.~\ref{sec:results} including relevant examples. We then consider in Sec.~\ref{sec:werner} a class of non-Gaussian states defined as mixtures of Gaussian states, which represent the CV counterparts to Werner states \cite{Mista_02b}. For these states, pseudospin measurements are found to be always more effective than Gaussian ones for steering detection. We finally draw our concluding remarks in Sec.~\ref{sec:concl}.

Overall, this paper represents a comprehensive exploration of EPR steering in CV systems beyond the Gaussian scenario, and may serve as an inspiration for further theoretical and experimental advances on the identification and exploitation of steering for quantum technologies.

\section{Preliminaries}\label{sec:prelim}
\subsection{Continuous variable systems and Gaussian states}\label{sec:gauss}

The object of our study is a CV quantum system, composed in general of  $N$ bosonic modes, and described by an infinite-dimensional Hilbert space constructed as a tensor product of the Fock spaces of each individual mode. The quadrature operators for a mode $j$ can be defined as $\hat{q}_j = (\hat{a}_j + \hat{a}^{\dagger}_j)/\sqrt{2}$, $\hat{p}_j = -i (\hat{a}_j - \hat{a}^\dagger_j)/ \sqrt{2}$, where $\hat{a}_j, \hat{a}_j^\dagger$ are the ladder operator satisfying $[\hat{a}_j, \hat{a}^\dagger_j]=1$, and $\hat{n}_j=\hat{a}^\dagger_j \hat{a}_j$ is the number operator, whose eigenvectors define the Fock basis, $\hat{n}_j \ket{n}_j = n_j \ket{n}_j$.
Collecting the quadrature operators for all the modes into a vector  $\hat{\boldsymbol{R}} = (\hat q_1, \hat p_1, \hat q_2, \hat p_2, \ldots, \hat q_N, \hat p_N)^{\sf T}$, the canonical commutation relations can be expressed as $[\hat R_j, \hat R_k] = i \left(\gr\omega^{\oplus N}\right)_{j,k}$ with $\gr\omega=i \boldsymbol{\sigma}^y$ and $\big\{\gr{\sigma}^x, \gr{\sigma}^y, \gr{\sigma}^z\big\}$ being the vector of Pauli matrices.

We will mainly focus our attention on Gaussian states \cite{ourreview,pirandolareview,ournewreview,serafozzi}, defined as those CV states whose Wigner phase space distribution is a multivariate Gaussian function of the form
\begin{equation}
\label{wigner}
W_\rho(\boldsymbol{\xi}) = \frac{1}{\pi^N \sqrt{\det{\V}}} \exp\big[-(\boldsymbol{\xi}-\boldsymbol{\delta})^{\sf T} \V^{-1} (\boldsymbol{\xi}-\boldsymbol{\delta})\big]\,,
\end{equation}
where $\boldsymbol{\xi} \in \mathbb{R}^{2N}$ denotes a phase space coordinate vector, $\boldsymbol{\delta} = \langle \hat{\boldsymbol{R}}\rangle$ is the displacement vector, and $\V$ is the covariance matrix collecting the second moments of the canonical operators,
\begin{equation}\label{covariancematrix}
 V_{j,k}=\langle\{\hat{R}_j-\delta_j,\hat{R}_k-\delta_k\}_+\rangle\,,
  \end{equation}
  with  $\{\cdot,\cdot\}_+$ standing for the anticommutator, and $\langle \cdot \rangle = \text{tr}\: [\hat{\rho} \ \cdot]$ denoting the expectation value.

  Since we are interested in correlations between the modes, we can assume without any loss of generality that the states have vanishing first moments, $\boldsymbol{\delta} = \boldsymbol 0$, as the latter can be adjusted by local displacements which have no effect on the correlations. The covariance matrix $\sig$ contains all the relevant information of a Gaussian state, and needs to obey the bona fide condition \cite{Simon94}
\begin{equation}\label{bona}
\V + i \boldsymbol{\omega}^{\oplus N} \geq 0\,,
\end{equation}
in order to correspond to a physical density matrix $\hat{\rho}$ in the Hilbert space.

In this work we will focus on a system of $N=2$ modes, $A$ and $B$, which can be accessed by two distant observers, respectively called Alice and Bob. Up to local unitaries, the covariance matrix of any two-mode Gaussian state can be written in the standard form \cite{Simon00,Duan00}
\begin{equation}\label{gamma}
\V\equiv \V_{AB}=\left(\begin{array}{c|c}
\gr\alpha & \gr\gamma \\ \hline
\gr\gamma\T  & \gr\beta
\end{array}\right) =
\left(\begin{array}{cc|cc}
a & & c & \\
& a & & d \\ \hline
c & & b & \\
& d & & b
\end{array}\right)\,.
\end{equation}
The real parameters $a,b,c,d$, constrained to inequality~(\ref{bona}), completely specify the global and marginal degrees of information and all forms of correlation in the state, and can be recast in terms of four local symplectic invariants of the covariance matrix \cite{extremal}. The states with $a=b$ are symmetric under swapping of the two modes. A particularly relevant subclass of Gaussian states is that of two-mode squeezed thermal (TMST) states, obtained from Eq.~(\ref{gamma}) by setting $d=-c$. These include as a special case of pure two-mode squeezed vacuum states, also known as EPR states, which are specified by
\begin{equation}
\label{gammaTMSV}
a=b=\cosh(2s)\,,\quad c=-d=\sinh(2s)\,,
\end{equation}
 in terms of a real squeezing parameter $s$.

\subsection{Steering criteria} \label{sec:steer}

In the quantum information language \cite{wiseman,wisepra}, EPR steering can be formalized in terms of entanglement verification when one of the parties is untrusted, or has uncharacterized devices. Suppose Alice wants to convince Bob, who does not trust her, that they are sharing an entangled state $\hat{\rho}$. Alice can then try and remotely prepare quantum ensembles on Bob's system that could not have been created without a shared entanglement in the first place. If she succeeds, with no need for any assumption on her devices, then entanglement is verified and the state shared by Alice and Bob is certified as  $A \rightarrow B$ steerable. In formula, a bipartite state $\hat{\rho}$ is $A \rightarrow B$ steerable if and only if the probabilities of all possible joint measurements cannot be factorized into a local hidden-variable/hidden-state form \cite{wiseman},
\begin{equation}\label{jointprob}
{\cal P} ( a,b|\hat{a},\hat{b},{{\hat{\rho}}} ) \neq \sum_{\lambda} {\cal P}_\lambda {\cal P} ( a|\hat{a},\lambda ) {\cal P} (b|\hat{b},{\hat{\rho}}_\lambda)\,, \quad \forall\, \hat{a}, \hat{b}\,,
\end{equation}
where $\lambda$ is a real variable, $\{\hat{\rho}_\lambda\}$ is an ensemble of marginal states for Bob, $\hat{a}, \hat{b}$ denote local observables for Alice and Bob, while $a$ and $b$ are their corresponding outcomes.

Detecting steerability from its definition (\ref{jointprob}) is challenging. To overcome this problem, several criteria have been developed to provide a more direct and experimentally friendly characterization of EPR steering \cite{steeringreview1,steeringreview2}. One such criterion, applicable to any (not necessarily Gaussian) CV bipartite state $\hat{\rho}$ with covariance matrix $\V$, yields that $\hat{\rho}$ is $A \rightarrow B$ steerable if \cite{wiseman}
\begin{equation}\label{wise}
\V + i (\boldsymbol{0}^{\oplus N_A} \oplus \boldsymbol{\omega}^{\oplus N_B}) \not\geq 0\,,
\end{equation}
where $N_A$ ($N_B$) is the number of modes of Alice's (Bob's) subsystem.
This corresponds to a steering test in which Alice and Bob perform Gaussian measurements, that is, they measure (linear combinations of) quadratures such as $\hat{q}_A, \hat{p}_A$ and $\hat{q}_B, \hat{p}_B$ respectively, by means of homodyne detections. Under the restriction of Gaussian measurements, inequality (\ref{wise}) is also necessary for $A \rightarrow B$ steerability if $\hat{\rho}$ is a Gaussian state. For two-mode states ($N_A=N_B=1$), the condition (\ref{wise}) can be rewritten simply as
\begin{equation}\label{wise2}
\det \gr\alpha > \det \V\,.
\end{equation}
In fact, for two-mode Gaussian states with covariance matrix in standard form as in Eq.~(\ref{gamma}), this necessary and sufficient condition is equivalent to the seminal variance criterion introduced in \cite{reid} to demonstrate the EPR paradox \cite{wisepra,steeringreview1,He2015,kogias,kogias2}.

As anticipated in the Introduction (Sec.~\ref{sec:intro}), however, Gaussian measurements are not always optimal to detect EPR steering of Gaussian states \cite{OneWayPryde,NhaSciRep}. More generally, one may need to resort to alternative criteria in order to witness EPR steering even in relatively simple states such as two-mode Gaussian states. A rather general approach to steering detection in bipartite states of any (finite or infinite) dimension was put forward in \cite{kogias3} in terms of a hierarchy of inequalities, constructed from the moment matrix corresponding to arbitrary pairs of measurements on Alice's and Bob's sides. As further detailed in the recent review \cite{steeringreview2}, this method is amenable to a numerical implementation via semidefinite programming, however it also provides an easily applicable analytical condition that is sufficient to reveal steering. Namely, a bipartite state $\hat{\rho}$ is $A \rightarrow B$ steerable if there exist spin-like measurement operators $\{\hat{s}^j_A\}$ and $\{\hat{t}^k_B\}$ for Alice and Bob, respectively, such that
\cite{kogias3}
\begin{equation}
\langle \hat{s}^x_A \otimes \hat{t}^x_B \rangle^2 +
\langle \hat{s}^y_A \otimes \hat{t}^y_B \rangle^2 +
\langle \hat{s}^z_A \otimes \hat{t}^z_B \rangle^2 > 1. \label{mom}
\end{equation}
In this work we adopt the criterion (\ref{mom}) to investigate EPR steering of two-mode Gaussian and non-Gaussian states by non-Gaussian pseudospin measurements, defined in the following.

\subsection{Pseudospin measurements} \label{sec:pseudo}

In this paper we consider two different sets of pseudospin measurements for CV systems.
Pseudospin operators of the first type were originally defined in \cite{chen} in order to investigate Bell nonlocality of EPR states. For a single mode (omitting the mode subscript for simplicity), they can be expressed as follows with respect to the Fock basis $\{\ket{n}\}$,
\begin{eqnarray}
\hat{S}^{x}&=&
\sum_{n=0}^{\infty} [\ket{2n}\bra{2n+1}+\ket{2n+1}\bra{2n}]\,,
 \nonumber \\
 \hat{S}^{y}&=&
\sum_{n=0}^{\infty} i [\ket{2n}\bra{2n+1}-\ket{2n+1}\bra{2n}]\,,\label{sxyz} \\
\hat{S}^{z}&=&
 \sum_{n=0}^{\infty} [\ket{2n+1}\bra{2n+1}-\ket{2n}\bra{2n}]=-\hat{P}\,, \nonumber
\end{eqnarray}
where $\hat{P}=(-1)^{\hat{n}}$ is the parity operator.
One can easily check that the operators $\{\hat{S}^j\}$ obey the standard SU(2) algebra just like the Pauli operators $\{\boldsymbol{\sigma}^j\}$, hence they can be regarded as infinite-dimensional analogues of the conventional spin observables, which motivates their denomination as pseudospin. Pseudospin operators as defined by Eq.~(\ref{sxyz}) have proven useful to analyze theoretically  bipartite and multipartite Bell nonlocality of Gaussian and non-Gaussian states \cite{chen,chen2,Mista_02,Mista_02b,Buqing}. However, evaluating expectation values of these operators requires handling the density matrix $\hat{\rho}$ expressed in the Fock basis, which may be quite nontrivial in general, as discussed in detail in Sec.~\ref{sec:fock}.

To sidestep this difficulty, an alternative set of pseudospin operators was introduced in \cite{gour1}. For a single mode, they can be expressed as follows in terms of even and odd superpositions of the eigenstates $\ket{q}$ of the position operator $\hat{q}$,
\begin{eqnarray}
\hat{\Pi}^{x}&=&
\int_{0}^{\infty} [\ket{\chi^+}\bra{\chi^-}+\ket{\chi^-}\bra{\chi^+}]dq\,,
 \nonumber \\
 \hat{\Pi}^{y}&=&
\int_{0}^{\infty} i [\ket{\chi^-}\bra{\chi^+}-\ket{\chi^+}\bra{\chi^-}]dq\,,\label{pxyz}
 \\
\hat{\Pi}^{z}&=&
\int_{0}^{\infty} [\ket{\chi^+}\bra{\chi^+}-\ket{\chi^-}\bra{\chi^-}]dq
=\hat{P}\,, \nonumber
\end{eqnarray}
where $\ket{\chi^{\pm}} = (\ket{q}\pm \ket{-q})/\sqrt{2}$. The operators $\{\hat{\Pi}^j\}$ also satisfy the standard SU(2) algebra, and will be referred to as pseudospin operators of the second type (or type-ii in short) in this paper, to distinguish them from the type-i ones of Eq.~(\ref{sxyz}). The type-ii pseudospin operators of Eq.~(\ref{pxyz}) admit a compact Wigner representation, given by \cite{gour2}
\begin{eqnarray}
W_{\Pi^x}(q,p) &=& \text{sgn}(q)\,,\quad  \nonumber \\
W_{\Pi^y}(q,p) &=& -\delta(q)\  \wp \frac{1}{p}\,, \quad  \label{wxyz}  \\
W_{\Pi^z}(q,p) &=& -\pi\ \delta(q)\ \delta(p), \nonumber
\end{eqnarray}
where $\wp$ denotes the principal value.
This allows one to evaluate expectation values of type-ii pseudospin operators directly from their Wigner function representation, with no need to resort to the Fock basis. Explicitly, for a two-mode state $\hat{\rho}$, we have \cite{gour2}
\begin{equation}
\label{typeii}
\langle \hat{\Pi}_A^j \otimes \hat{\Pi}^k_B\rangle = \frac{1}{(2\pi)^2} \int d^4 \boldsymbol{\xi}\ W_\rho(\boldsymbol{\xi})\ W_{\Pi_A^j}(q_A,p_A)\ W_{\Pi_B^k}(q_B,p_B)\,,
\end{equation}
with $\boldsymbol{\xi}=(q_A, p_A, q_B, p_B)\T$.
The type-ii pseudospin operators have also been employed for studies of bipartite and tripartite Bell nonlocality \cite{gour1,gour2,ferraro23}.

A comparison between the performance of the two types of pseudospin operators for verifying the quantumness of correlations in a model of the early universe was also recently reported \cite{vennin}. However, both type-i and type-ii pseudospin measurements remain challenging to implement experimentally with current technology.

\section{Fock representation of two-mode squeezed thermal states}\label{sec:fock}

In this Section we obtain a result of importance in its own right, that is, we derive an explicit expression for the elements of the density matrix of an arbitrary Gaussian TMST state $\hat{\rho}$, with vanishing first moments and covariance matrix  $\V$ in standard form given by Eq.~(\ref{gamma}) with $d=-c$, in the Fock basis $\{\ket{mn}_{AB} \equiv \ket{m}_A \ket{n}_B\}_{m,n=0,1,\ldots}$.

\begin{figure}[t]
\begin{center}
\includegraphics[width=8.8cm]{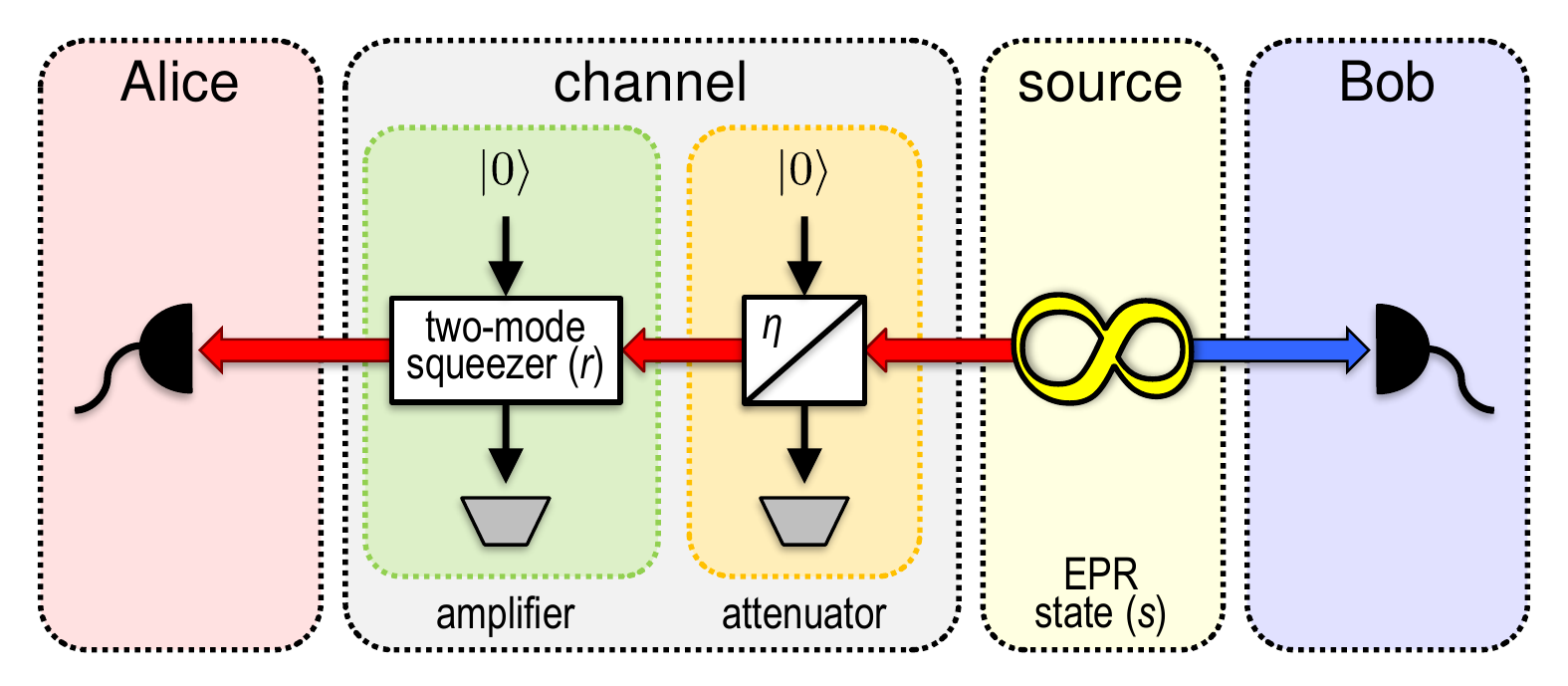}
\protect\caption{(Color online) Scheme to generate an arbitrary TMST state \cite{Pirandola_14}. An EPR source (yellow box) prepares a two-mode squeezed
vacuum state with squeezing parameter $s$, Eq.~(\ref{rhoTMSV}). One mode (blue arrow) of the state freely propagates towards Bob while
the other mode (red arrow) propagates through a phase-insensitive Gaussian channel (light gray box) towards Alice. The channel can be decomposed  \cite{Caruso_06,Garcia-Patron_12} into a quantum-limited attenuator realized by a beam splitter with intensity transmissivity $\eta$
(orange box) followed by a quantum-limited amplifier implemented by a two-mode squeezer with a squeezing parameter $r$ (green box).
The state shared by Alice and Bob is then a TMST state $\hat{\rho}^{\text{TMST}}_{AB}(s,\eta,r)$ with covariance matrix given by Eq.~(\ref{gammaTMST2}). See text for further details. \label{trick}}
\end{center}
\end{figure}

For this purpose we use the trick \cite{Pirandola_14} depicted in Fig.~\ref{trick} that any TMST state can be prepared by applying a
suitable single-mode phase-insensitive Gaussian channel $\mathcal{E}$ to mode $A$ of a two-mode squeezed vacuum state
\begin{equation}\label{rhoTMSV}
\hat{\rho}^{\text{EPR}}_{AB}(s)= \ket{\psi_s}_{AB}\bra{\psi_s},\ \ \ \ket{\psi_s}_{AB}=\frac{1}{\cosh(s)}\sum_{m=0}^{\infty}\tanh^m(s)\ket{mm}_{AB}\,,
\end{equation}
where  $s$  is the squeezing parameter, which possesses a covariance matrix $\V_{AB}^{\text{EPR}}(s)$ given  by Eqs.~(\ref{gamma})--(\ref{gammaTMSV}).
Next, we use the fact that any single-mode phase-insensitive Gaussian channel $\mathcal{E}$ can be
decomposed into a sequence of a quantum-limited attenuator (i.e., pure-loss) channel $\mathcal{L}_\eta$ with transmissivity $\eta$ followed by a
quantum-limited amplifier $\mathcal{A}_r$ with gain $\cosh^{2}(r)$ \cite{Caruso_06,Garcia-Patron_12}, i.e., $\mathcal{E}=\mathcal{A}_r\circ\mathcal{L}_\eta$.

On the level of covariance matrix, the pure-loss channel
$\mathcal{L}_\eta$ on mode $A$ can be implemented by mixing the mode with an ancillary mode $A'$, initially in the vacuum state with covariance matrix $\V_{A'}=\openone$,
on a beam splitter described by the symplectic matrix\footnote{A unitary operation ${\hat U}$ which maps Gaussian states into Gaussian states can be described by a symplectic transformation $S$ which acts by congruence on covariance matrices, $\V \mapsto S \V S\T$.}
\begin{equation}\label{SBS}
S^{\text{BS}}_\eta=\left(\begin{array}{cc}
\sqrt{\eta}\openone & -\sqrt{1-\eta}\openone \\
\sqrt{1-\eta}\openone  & \sqrt{\eta}\openone
\end{array}\right)\,.
\end{equation}
By taking the covariance matrix $\V^{\text{EPR}}_{AB}(s)\oplus\V_{A'}$ of the three modes $A,B$ and $A'$,
transforming modes $A$ and $A'$ via the symplectic matrix (\ref{SBS}), and dropping mode $A'$, we get the covariance matrix of the output state of modes $A$ and $B$ after the pure-loss channel,
\begin{equation}\label{gammaABprimed}
\mathcal{L}_\eta[\V^{\text{EPR}}_{AB}(s)]=\left(\begin{array}{cc}
(b\eta+1-\eta)\openone & \sqrt{\eta(b^2-1)}\gr\sigma^z \\
\sqrt{\eta(b^2-1)}\gr\sigma^z  & b\openone
\end{array}\right)\,,
\end{equation}
with $b=\cosh(2s)$.
Likewise, the amplifier $\mathcal{A}_r$ can be realized by mixing mode $A$ with another vacuum ancillary mode $A''$ in a two-mode
squeezer with squeezing parameter $r$, described by the symplectic matrix
\begin{equation}\label{STMS}
S^{\text{TM}}_r=\left(\begin{array}{cc}
\cosh(r)\openone & \sinh(r)\gr\sigma^{z} \\
\sinh(r)\gr\sigma^{z}  & \cosh(r)\openone
\end{array}\right)\,.
\end{equation}
 By transforming modes $A$ and $A''$ of the intermediate state with covariance matrix (\ref{gammaABprimed}) via the symplectic matrix (\ref{STMS}), and dropping mode $A''$, we finally get the output covariance matrix of modes $A$ and $B$ (see Fig.~\ref{trick}), given by \cite{Pirandola_14}
\begin{equation}\label{gammaTMST2}
\begin{split}
\V^{\text{TMST}}_{AB}(s,\eta,r)&=(\mathcal{A}_r \circ \mathcal{L}_\eta)[\V^{\text{EPR}}_{AB}(s)]\\
&=\left(\begin{array}{cc}
(\tau b+\zeta)\openone & \sqrt{\tau(b^2-1)}\gr\sigma^{z} \\
\sqrt{\tau(b^2-1)}\gr\sigma^{z}  & b\openone
\end{array}\right),
\end{split}
\end{equation}
with $\tau=\eta\cosh^{2}(r)$, $\zeta=(1-\eta)\cosh^{2}(r)+\sinh^{2}(r)$.
In other words, Eq.~(\ref{gammaTMST2}) is the covariance matrix of a TMST state $\hat{\rho}^{\text{TMST}}_{AB}$ in standard form, given by Eq.~(\ref{gamma}) with
\begin{eqnarray}
a&=&\eta\cosh^{2}(r) \cosh(2s)+(1-\eta)\cosh^{2}(r)+\sinh^{2}(r)\,,\nonumber\\
b&=&\cosh(2s)\,, \label{ac}\\
-d&=&c=\sqrt{\eta}\cosh(r)\sinh(2s)\,.  \nonumber
\end{eqnarray}
Since for any admissible values of parameters $a,b$ and $c=-d$ there is always
a physical channel $\mathcal{E}$ for which the covariance matrices  (\ref{gammaTMST2}) and (\ref{gamma}) coincide, we can always
parameterize the standard form covariance matrix of a TMST state as in Eq.~(\ref{gammaTMST2}).

Let us now move to the evaluation of the Fock basis elements of the TMST state, exploiting the parametrization of Fig.~\ref{trick}.
By applying the channel $\mathcal{E}$ to the
first mode of the density matrix (\ref{rhoTMSV}) and using the decomposition $\mathcal{E}=\mathcal{A}_r\circ\mathcal{L}_\eta$, the matrix element to be
evaluated boils down to
\begin{equation}\label{Fock}
\langle m_{1}m_{2}|\hat{\rho}^{\text{TMST}}_{AB}|n_{1}n_{2}\rangle=(1-\varsigma^{2})\varsigma^{m_{2}+n_{2}}\langle m_{1}|(\mathcal{A}_r\circ\mathcal{L}_\eta)
(|m_{2}\rangle\langle n_{2}|)|n_{1}\rangle,
\end{equation}
where we set $\varsigma=\tanh(s)$ and used the linearity of the channel $\mathcal{E}$.

First, we need to calculate how the pure-loss channel $\mathcal{L}_\eta$ transforms the operator $|m\rangle\langle n|$. On the state vector level, a beam splitter unitary ${\hat U}^{\text{BS}}_\eta$ with symplectic matrix (\ref{SBS}) transforms the state $|m\rangle_{A}|0\rangle_{A'}$ as
\begin{eqnarray}\label{UBSm0}
|m\rangle_{A}|0\rangle_{A'}&\rightarrow&|\phi\rangle_{AA'}\equiv {\hat U}^{\text{BS}}_\eta |m\rangle_{A}|0\rangle_{A'}   \\
&=&{\hat U}^{\text{BS}}_\eta\frac{(\hat{a}_A^{\dag})^{m}}{\sqrt{m!}}{{\hat{U}^{\dag\text{BS}}_\eta}}|0\rangle_{A}|0\rangle_{A'}\nonumber\\
&=&\sum_{k=0}^m  \sqrt{\binom{m}{k} \frac{\eta^{{k}}} {(1-\eta)^{{k-m}}}}   \ket{k,m-k}_{AA'}, \nonumber
\end{eqnarray}
where we used the relation ${\hat U}^{\text{BS}}_\eta|0\rangle_{A}|0\rangle_{A'}=|0\rangle_{A}|0\rangle_{A'}$, the
transformation rule $\hat{U}_{\eta}^{BS}\hat{a}_{A}^{\dag}\hat{U}_{\eta}^{\dag BS}=
\sqrt{\eta}\hat{a}_{A}^{\dag}+\sqrt{1-\eta}\hat{a}_{A'}^{\dag}$, and the binomial theorem.
Hence, if we trace out the ancilla $A'$ from the state $|\phi\rangle_{AA'}$ of Eq.~(\ref{UBSm0}), we get the sought expression
\begin{eqnarray}\label{L}
\mathcal{L}_\eta(|m\rangle\langle n|)=\eta^{\frac{m+n}{2}}\sum_{k=0}^{\mathrm{min}\{m,n\}}\sqrt{{m\choose k}{n\choose k}}(\eta^{-1}-1)^{k}
|m-k\rangle\langle n-k|.\nonumber\\
\end{eqnarray}

We now need to calculate how the amplifier $\mathcal{A}_r$ acts
on the operator $|m\rangle\langle n|$. Similarly to the previous case, a two-mode squeezer unitary ${\hat U}^{\text{TM}}_r$  with
symplectic matrix (\ref{STMS}) transforms the state $|m\rangle_{A}|0\rangle_{A''}$ as
\begin{eqnarray}\label{UTMSm0}
|m\rangle_{A}|0\rangle_{A''}&\rightarrow&|\varphi\rangle_{AA''}\equiv {{\hat{U}^{\text{TM}}_r}}|m\rangle_{A}|0\rangle_{A''}  \\
&=&{{\hat{U}^{\text{TM}}_r}}\frac{(\hat{a}_A^{\dag})^{m}}{\sqrt{m!}}{{\hat{U}^{\dag\text{TM}}_r}}{{\hat{U}^{\text{TM}}_r}}|0\rangle_{A}|0\rangle_{A''}\nonumber\\
&=&\frac{1}{[\cosh(r)]^{m+1}}\sum_{l=0}^{\infty}\sqrt{{m+l\choose m}}\tanh^l(r)|l+m,l\rangle_{AA''}, \nonumber
\end{eqnarray}
where we used the relations ${{\hat{U}^{\text{TM}}_r}} |0\rangle_{A}|0\rangle_{A''} = \ket{\psi_r}_{AA''}$ with $\ket{\psi_r}$ defined in Eq.~(\ref{rhoTMSV}),
${{\hat{U}^{\text{TM}}_r}}\hat{a}_A^{\dag}{{\hat{U}^{\dag\text{TM}}_r}}=\cosh(r)\hat{a}_A^{\dag}-\sinh(r)\hat{a}_{A''}$,
and the binomial theorem.
Now, by tracing out the ancilla $A''$ from the state $|\varphi\rangle_{AA''}$ of Eq.~(\ref{UTMSm0}),
we find that the amplifier transforms the operator $|m\rangle\langle n|$ in the following way
\begin{eqnarray}\label{A}
\mathcal{A}_r(|m\rangle\langle n|)&=&\frac{1}{[\cosh(r)]^{m+n+2}}\sum_{l=0}^{\infty}\sqrt{{m+l\choose m}{n+l\choose n}}\left[\tanh(r)\right]^{2l}\nonumber\\
&&\times\ |m+l\rangle\langle n+l|\,.
\end{eqnarray}

Having formulae (\ref{L}) and (\ref{A}) in hands we are now in the position to calculate the matrix elements
(\ref{Fock}) of a TMST state in Fock basis. A rather lengthy algebra finally yields the main result of this Section:
\begin{eqnarray}\label{Fockfinal}
&&\!\!\!\!\!\!\!\!\langle m_{1}m_{2}|\hat{\rho}^{\text{TMST}}_{AB}|n_{1}n_{2}\rangle   \\
&&=\delta_{m_1+n_2,n_{1}+m_{2}}\frac{(1-\varsigma^{2})}{\cosh^{2}(r)}\left[\frac{\varsigma \sqrt{\eta}}{\cosh(r)}\right]^{m_{2}+n_{2}}\left[\tanh(r)\right]^{2(m_{1}-m_{2})} \nonumber \\ &&\ \ \ \times
\sum_{k=\mathrm{max}\{0,m_{2}-m_{1}\}}^{\mathrm{min}\{m_{2},n_{2}\}}\sqrt{{m_{2}\choose k}{n_{2}\choose k}{m_{1}\choose m_{2}-k}{n_{1}\choose n_{2}-k}}\nonumber\\
&&\ \ \ \times\left[\sqrt{\frac{1-\eta}{\eta}}\sinh(r)\right]^{2k}.\nonumber
\end{eqnarray}
This formula, which to the best of our knowledge has not appeared elsewhere, provides the exact Fock basis representation for the most relevant class of two-mode Gaussian states, encompassing and generalizing previously known special cases such as the instances where only one of the channels acts on mode $A$, either the pure-loss ${\cal L}_\eta$ (i.e., $r=0$) or the amplifier ${\cal A}_r$ (i.e., $\eta=1$) \cite{Mista_02,NhaSciRep,Mista_14}. Note that, due to the presence of the Kronecker symbol, the matrix elements in Eq.~(\ref{Fockfinal}) vanish if  $m_{1}+m_{2}+n_{1}+n_{2}$ is odd, as it should be \cite{Mista_14}. In the \hyperref[sec:app]{Appendix}, we explore further applications of Eq.~(\ref{Fockfinal}) to derive compact expressions for multidimensional Hermite polynomials.

For completeness, we also report the explicit state vectors of all the modes involved in the scheme of Fig.~\ref{trick} before tracing out the ancillae.
Given an initial two-mode squeezed state of the system modes $A$ and $B$ as in  Eq.~(\ref{rhoTMSV}) with squeezing $s$ (EPR source), the state of modes $A,B$ and $A'$ after the action of the beam splitter with transmissivity $\eta$ on $A$ and $A'$ can be written using Eq.~(\ref{UBSm0}) as
\begin{eqnarray}\label{lossA}
&&\!\!\!\!\!\!\!\!\!\!\!\!\! \ket{\Phi_{s,\eta}}_{ABA'}=
 ({\hat U}^{\text{BS}}_{\eta\ {AA'}} \otimes \openone_B)(\ket{\psi_s}_{AB} \otimes \ket{0}_{A'})    \\
  &=& \sqrt{1-\varsigma^2} \sum_{m=0}^\infty \varsigma^m \sum_{k=0}^m  \sqrt{\binom{m}{k} \frac{\eta^{{k}}} {(1-\eta)^{{k-m}}}}  \ket{k,m,m-k}_{ABA'}.  \nonumber
\end{eqnarray}

The final state of all modes $A,B,A',A''$ after the successive action of the two-mode squeezer with squeezing $r$ on $A$ and $A''$ can then be written using Eq.~(\ref{UTMSm0}) as
\begin{eqnarray}\label{TMSTfock}
&&\!\!\!\!\!\!\!\!\! \ket{\Psi_{s,\eta,r}}_{ABA'A''}=
 ({\hat U}^{\text{TM}}_{r\ {AA''}} \otimes \openone_{BA'})(\ket{\Phi_{s,\eta}}_{ABA'} \otimes \ket{0}_{A''})  \\
  &&=
  \sqrt{1-\varsigma^2} \sum_{m=0}^\infty \varsigma^m \sum_{k=0}^m  \sqrt{\binom{m}{k} \frac{\eta^{{k}}} {(1-\eta)^{{k-m}}}}  \frac{1}{[\cosh(r)]^{k+1}}  \nonumber\\
&&\ \ \ \times \sum_{l=0}^\infty \sqrt{\binom{k+l}{k}} [\tanh(r)]^{l}\ket{k+l,m,m-k,l}_{ABA'A''}. \nonumber
\end{eqnarray}

By tracing over the ancillary modes, one recovers the TMST state $\text{tr}_{A'A''}\ket{\Psi_{s,\eta,r}}_{ABA'A''}\bra{\Psi_{s,\eta,r}} \equiv \hat{\rho}^{\text{TMST}}_{AB}(s,\eta,r)$ described by Eq.~(\ref{ac}) on the level of covariance matrix and by Eq.~(\ref{Fockfinal}) on the level of Fock space.

\section{Steering of two-mode Gaussian states: Gaussian versus non-Gaussian measurements}\label{sec:results}

We now investigate EPR steering of two-mode Gaussian states as detected by the criteria presented in Section~\ref{sec:steer}, looking especially for instances in which  superiority of non-Gaussian measurements over Gaussian ones can be recognized. We focus in particular on the criterion of Eq.~(\ref{mom}) \cite{kogias3} evaluated using pseudospin measurements as reported in Sec.~\ref{sec:pseudo}.

\subsection{Expectation values of pseudospin measurements}

\subsubsection{Type-i}

Thanks to the results of Sec.~\ref{sec:fock}, namely Eq.~(\ref{Fockfinal}), we are able to evaluate the expectation values of type-i pseudospin operators, defined by Eq.~(\ref{sxyz}) \cite{chen}, for a general TMST $\hat{\rho}^{\text{TMST}}_{AB}(s,\eta,r)$ parameterized by initial squeezing $s$, attenuator transmissivity $\eta$, and amplifier squeezing $r$, according to the scheme of Fig.~\ref{trick}.
After some algebra, we find
\begin{eqnarray}
\langle \hat{S}^x_A \otimes \hat{S}^x_B \rangle&=&\frac{2}{\cosh^2(s)} \sum_{n,l=0}^\infty [\tanh(s)]^{4n+1} \left[\Gamma_{nl}(\eta,r)\! +\! \Upsilon_{nl}(\eta,r)\right], \nonumber \\
\langle \hat{S}^y_A \otimes \hat{S}^y_B \rangle&=&-\langle \hat{S}^x_A \otimes \hat{S}^x_B \rangle, \label{sxyztmst} \\
\langle \hat{S}^z_A \otimes \hat{S}^z_B \rangle&=&\langle \hat{\rho}^{\text{TMST}}_{AB}(s,\eta,r) \rangle = \frac{1}{\sqrt{\det \V^{\text{TMST}}_{AB}(s,\eta,r)}}, \nonumber
\end{eqnarray}
with
\begin{eqnarray*}
\Gamma_{nl}(\eta,r)&=&\sum_{k=0}^n \eta ^{2k+\frac12} (1-\eta )^{2n-2k} [\cosh(r)]^{-(4 k+3)} [\tanh(r)]^{4 l} \\
&& \ \ \ \ \times  \sqrt{\binom{2 k+2 l}{2 k} \binom{2 k+2 l+1}{2 k+1} \binom{2 n}{2 k} \binom{2 n+1}{2 k+1}}, \\
\Upsilon_{nl}(\eta,r)&=&\sum_{k=0}^{n-1} \eta ^{2 k+\frac32} (1-\eta )^{2n-2k-1} [\cosh(r)]^{-(4 k+5)} [\tanh(r)]^{4 l+2} \\
 && \ \ \ \ \times \sqrt{\binom{2 k+2 l+2}{2 k+1} \binom{2 k+2 l+3}{2 k+2} \binom{2 n}{2 k+1} \binom{2 n+1}{2 k+2}}.
\end{eqnarray*}
The expressions in Eq.~(\ref{sxyztmst}) can be evaluated numerically by truncating the sums over $n$ and $l$ to an appropriately large integer depending on the value of the parameters $s,\eta,r$.

Before going further, let us note that the steering inequality (\ref{mom}) for type-i pseudospin operators
(\ref{sxyz}) can be interpreted in the context of mapping of CV modes onto qubits \cite{Mista_02,Kraus_04,Paternostro_04,Paternostro_10}.
Namely, it is possible to map via a nonliner Jaynes-Cummings interaction a density matrix $\hat{\rho}_{A}$
of a single mode $A$ onto a density matrix $\hat{\rho}_{1}$ of a single qubit $1$, such that
$\langle \hat{S}^{j}\rangle_{\hat{\rho}_{A}}\equiv\text{tr}
[\hat{S}^{j}\hat{\rho}_{A}]=\text{tr}[\gr{\sigma}^{j}\hat{\rho}_{1}]\equiv\langle\gr{\sigma}^{j}\rangle_{\hat{\rho}_{1}}$  \cite{Mista_02}.
Similarly, if we map locally modes $A$ and $B$ of a two-mode state $\hat{\rho}_{AB}$ onto two qubits $1$ and $2$,
the qubits will end up in a state $\hat{\rho}_{12}$ for which
$\langle \hat{S}_{A}^{i}\otimes\hat{S}_{B}^{j}\rangle_{\hat{\rho}_{AB}}=
\langle \gr{\sigma}^{i}\otimes\gr{\sigma}^{j}\rangle_{\hat{\rho}_{12}}$. Thus the analysis of EPR
steering for a two-mode state using type-i pseudospin measurements can be seen as a mapping onto a two-qubit state followed by the analysis of EPR
steering for the two-qubit state using conventional spin measurements. Since the mapping does not preserve entanglement \cite{Mista_02b,Paternostro_04, Paternostro_10,Passing_09}, i.e.,
some entangled two-mode states are mapped onto separable two-qubit states, we may expect that the same
holds true also for steering, that is, a two-mode state can be steerable although the corresponding two-qubit
state is unsteerable.

\subsubsection{Type-ii}

In the case of type-ii pseudospin operators, defined by Eq.~(\ref{pxyz}) \cite{gour1,gour2}, we can evaluate analytically their expectation values for all two-mode Gaussian states, specified in general by a standard form covariance matrix $\V_{AB}$ as in Eq.~(\ref{gamma}) as a function of $a,b,c,d$. Exploiting the formulation in terms of Wigner functions, Eq.~(\ref{typeii}), we get
\begin{eqnarray}
\langle \hat{\Pi}^x_A \otimes \hat{\Pi}^x_B \rangle&=&\frac{2}{\pi}\arctan\sqrt{\frac{c^2}{ab-c^2}}, \nonumber \\
\langle \hat{\Pi}^y_A \otimes \hat{\Pi}^y_B \rangle&=&\frac{2}{\pi\sqrt{\det \V_{AB}}}\arctan\sqrt{\frac{d^2}{ab-d^2}}, \label{pxyzabcd} \\
\langle \hat{\Pi}^z_A \otimes \hat{\Pi}^z_B \rangle&=& \frac{1}{\sqrt{\det \V_{AB}}}, \nonumber
\end{eqnarray}
where to obtain the second equation we resorted to Parseval's theorem.

\subsection{Steering analysis, examples and discussion}

Let us denote the combination of moments in the left-hand side of the EPR steering criterion Eq.~(\ref{mom}) evaluated on a state $\hat{\rho}_{AB}$ as $M_{\hat{\rho}_{AB}}^{(j)}$, with $j=\text{i},\text{ii}$ denoting the pseudospin type. Explicitly,
\begin{equation}
\label{mammt}
\begin{split}
M_{\hat{\rho}_{AB}}^{(\text{i})} &= \langle \hat{S}^x_A \otimes \hat{S}^x_B \rangle^2 + \langle \hat{S}^y_A \otimes \hat{S}^y_B \rangle^2 + \langle \hat{S}^z_A \otimes \hat{S}^z_B \rangle^2\,, \\
M_{\hat{\rho}_{AB}}^{(\text{ii})} &= \langle \hat{\Pi}^x_A \otimes \hat{\Pi}^x_B \rangle^2 + \langle \hat{\Pi}^y_A \otimes \hat{\Pi}^y_B \rangle^2 + \langle \hat{\Pi}^z_A \otimes \hat{\Pi}^z_B \rangle^2\,.
\end{split}
\end{equation}

\subsubsection{Two-mode squeezed vacuum states}
\begin{figure}[t]
\begin{center}
\includegraphics[width=8cm]{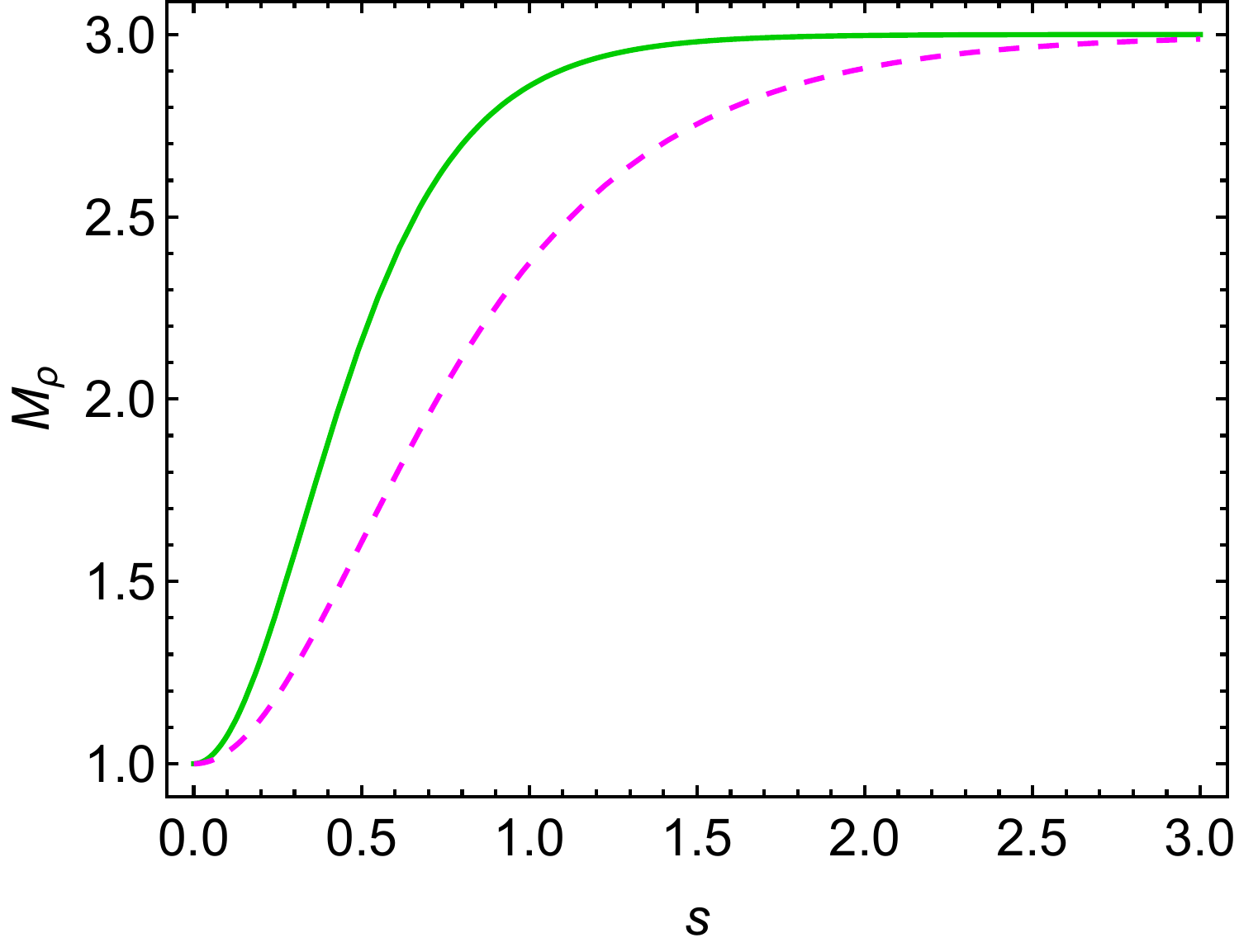}
\protect\caption{(Color online) Plot of the left-hand side $M_{\hat{\rho}^{\text{EPR}}_{AB}}^{(j)}(s)$ [Eq.~(\ref{mTMSV})] of the squeezing criterion [Eq.~(\ref{mom})] using pseudospin measurements of type $j$, with $j=\text{i}$ (solid green) and $j=\text{ii}$ (dashed magenta), for a two-mode squeezed vacuum state $\hat{\rho}^{\text{EPR}}_{AB}(s)$ [Eq.~(\ref{rhoTMSV})] as a function of the squeezing parameter $s$. All the quantities plotted are dimensionless. \label{onevstwo}}
\end{center}
\end{figure}

We begin by comparing the two types of pseudospin measurements on a two-mode squeezed vacuum state $\hat{\rho}^{\text{EPR}}_{AB}(s)$, defined by Eq.~(\ref{rhoTMSV}) or equivalently by its covariance matrix $\V_{AB}^{\text{EPR}}(s)$ with elements given in Eq.~(\ref{gammaTMSV}). Referring to the scheme of Fig.~\ref{trick}, this amounts to a pure EPR source without any further noisy channel on $A$ (i.e., $\eta=1, r=0)$. It is well known that this state is entangled, EPR steerable and Bell nonlocal as soon as $s>0$. It thus presents itself as an easy testground for the criterion of Eq.~(\ref{mom}).

The expressions in Eq.~(\ref{mammt}) are straightforward to compute for the two-mode squeezed state, yielding
\begin{equation}
\label{mTMSV}
M_{\hat{\rho}^{\text{EPR}}_{AB}}^{(\text{i})}(s) = 1+2\tanh^2(2s)\,,\quad
M_{\hat{\rho}^{\text{EPR}}_{AB}}^{(\text{ii})}(s) = 1+(8/\pi^2)\, \text{gd}^2(2s)\,,
\end{equation}
where $\text{gd}(z)=2 \arctan (e^z)-\pi/2$ is the Gudermannian function. As plotted in Fig.~\ref{onevstwo}, both quantities in Eq.~(\ref{mTMSV}) are larger than $1$ for any $s>0$, and increase monotonically as a function of $s$ reaching their maximum value of $3$ in the limit $s \rightarrow \infty$. This confirms that the criterion of Eq.~(\ref{mom}) is able to reveal maximum steerability of the EPR state using either type of pseudospin measurements, in agreement with previous studies of Bell nonlocality \cite{chen,gour1,gour2}. However, we also notice that $M_{\hat{\rho}^{\text{EPR}}_{AB}}^{(\text{i})}(s) \geq M_{\hat{\rho}^{\text{EPR}}_{AB}}^{(\text{ii})}(s)$ in general, which suggests that type-ii observables are less sensitive than type-i ones for steering detection using the adopted criterion. This will be confirmed for more general states in the following.

\subsubsection{Two-mode squeezed states with loss on Alice}

Next, we consider an important example of noisy Gaussian state, namely a TMST state resulting from the action of a pure-loss channel ${\cal L}_\eta$ with transmissivity $\eta$ on mode $A$ of an EPR state $\hat{\rho}^{\text{EPR}}_{AB}(s)$. This state, that will be denoted by $\hat{\rho}^{\text{TMST}}_{AB}(s,\eta,0)$ according to the notation of Sec.~\ref{sec:fock} (see Fig.~\ref{trick}), arises naturally in quantum key distribution, where  ${\cal L}_\eta$ models attenuation due to transmission losses. According to Eq.~(\ref{wise2}), the state $\hat{\rho}^{\text{TMST}}_{AB}(s,\eta,0)$ is $A \rightarrow B$ steerable by Gaussian measurements if and only if  \cite{wiseman}
\begin{equation}\label{etag}
\eta > \frac12\,.
\end{equation}
However, recently Refs.~\cite{OneWayPryde,NhaSciRep} found that Eq.~(\ref{etag}) is not a critical threshold for one-way steerability, as the state $\hat{\rho}^{\text{TMST}}_{AB}(s,\eta,0)$  can be steered from Alice to Bob even at lower values of $\eta$ if using suitable non-Gaussian measurements. Before presenting the results of our analysis, let us provide a bit more details on the findings of \cite{OneWayPryde,NhaSciRep}.

The authors of \cite{OneWayPryde} considered an equatorial family of type-i pseudospin measurements with $\hat{S}^{\theta}=\cos{(\theta)}\hat{S}^x+\sin{(\theta)}\hat{S}^y$ and applied it together with $\hat{S}^z$ to the following nonlinear steering criterion \cite{HowardCri}
\begin{equation}
\int_{-\pi}^{\pi} d\theta~\langle \hat{A}^{\theta} \hat{S}^{\theta}\rangle > \frac{2}{\pi} \Bigg(P_+ \sqrt{1-Z_+^2}+P_- \sqrt{1-Z_-^2}\Bigg)\,, \label{howard}
\end{equation}
where the measurement $\hat{A}^{\theta}$ that Alice performs on her mode is informed by Bob's choice of  $\hat{S}^{\theta}$, and $P_{\pm}$ are the probabilities that Alice obtains results $\pm1$ for her observable $\hat{A}^z$, while $Z_{\pm}$ are Bob's respective conditional expectation values. The fulfillment of Eq.~(\ref{howard}) implies that steering can be demonstrated from Alice to Bob.

Instead of pseudospin measurements, the authors in Ref.~\cite{NhaSciRep} defined a collection of $n^2$ orthogonal observables $\{\hat{A}^{(n)}\}=\{\lambda_k, \lambda_{kl}^{\pm}\}$ $(k,l=0,1,\dots, n-1)$ where
\begin{equation}
\lambda_k =
\ket{k}\bra{k},~~~~~\lambda_{kl}^{\pm}=
\frac{\ket{k}\bra{l}\pm\ket{l}\bra{k}}{\sqrt{2}\ (-1)^{\frac14 \mp \frac14}}~~~(k<l)\,. \label{Nha}
\end{equation}
If the correlation matrix $C_{nn'}$ of a two-mode state $\rho_{AB}$, with elements $(C_{nn'})_{ij}\equiv \langle \hat{A}_{i}^{(n)} \otimes \hat{B}_j^{(n')}\rangle-\langle \hat{A}_{i}^{(n)}\rangle \langle \hat{B}_j^{(n')}\rangle$, violates the local uncertainty relations of these non-Gaussian measurements \cite{NhaSciRep}, i.e., if
\begin{equation}
\parallel C_{nn'}\parallel_{\text{tr}}\  > \sqrt{\bigg(n\langle \openone_{n}^{A}\rangle-\sum_{j=1}^n\langle \hat{A}_{j}^{(n)}\rangle^2 \bigg)\bigg(\langle \openone_{n'}^{B}\rangle-\sum_{j=1}^{n'}\langle \hat{B}_{j}^{(n')}\rangle^2 \bigg)}\,,
 \label{NhaCri}
\end{equation}
then $\rho_{AB}$ is steerable from Alice to Bob.

\begin{figure}[t]
\begin{center}
\includegraphics[width=8cm]{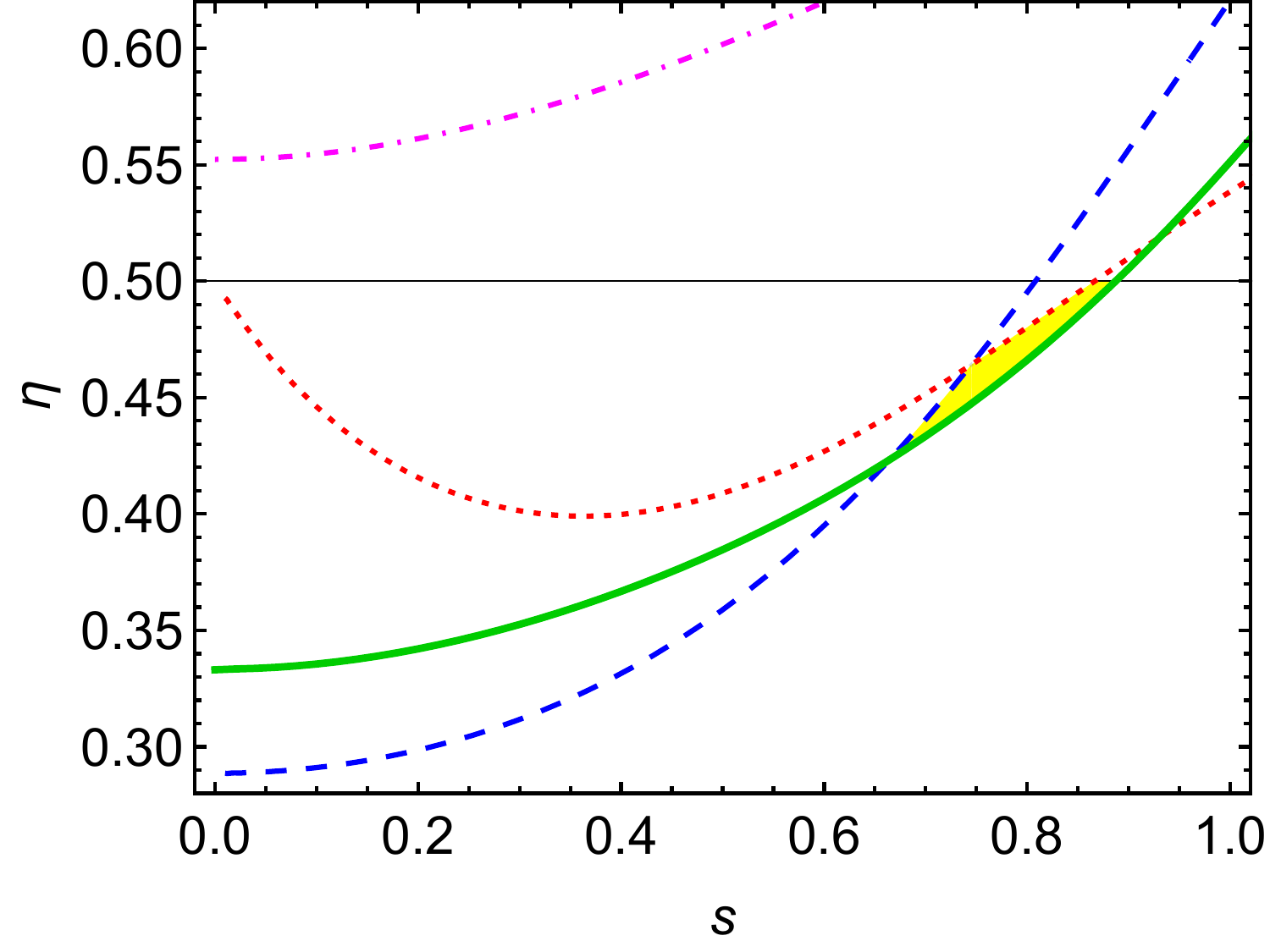}
\protect\caption{(Color online) Comparison of EPR steering criteria for the lossy state $\hat{\rho}^{\text{TMST}}_{AB}(s,\eta,0)$ as a function of the initial squeezing $s$ and the attenuator transmissivity $\eta$ (see Fig.~\ref{trick}). $A \rightarrow B$ steering is detected by each criterion in the parameter region above the corresponding threshold curve. The considered criteria are: steerability by Gaussian measurements; Eq.~(\ref{etag}) \cite{wiseman} (thin solid black), nonlinear steering criterion with type-i pseudospin measurements, Eq.~(\ref{howard}) \cite{OneWayPryde} (dashed blue); local uncertainty relation criterion with two-level orthogonal observables, Eq.~(\ref{NhaCri}) \cite{NhaSciRep} (dashed red); moment criterion, Eq.~(\ref{mom}) \cite{kogias3}, evaluated in this paper with type-i pseudospin measurements \cite{chen} (thick solid green); moment criterion, Eq.~(\ref{mom}) \cite{kogias3}, evaluated in this paper with type-ii pseudospin measurements \cite{gour1} (dot-dashed magenta). In the shaded yellow region of parameters, steering is identified only by our criterion based on type-i pseudospin observables. All the quantities plotted are dimensionless. \label{lossa}}
\end{center}
\end{figure}

To further investigate the EPR steering of the lossy state $\hat{\rho}^{\text{TMST}}_{AB}(s,\eta,0)$ by non-Gaussian measurements, we have evaluated the criterion of Eq.~(\ref{mom}) based on pseudospin measurements. Instead of reporting the explicit expressions for the quantities of Eq.~(\ref{mammt}), we plot in Fig.~\ref{lossa} the threshold curves such that $M_{\hat{\rho}^{\text{TMST}}_{AB}}^{(j)}(s,\eta,0) = 1$, for $j=\text{i}$ (thick solid green) and $j=\text{ii}$ (dot-dashed magenta), in the space of parameters $(s,\eta)$. Steering from Alice to Bob according to the chosen measurements is demonstrated in the region above the corresponding threshold curve. The figure also compares our findings with the thresholds arising from Gaussian measurements [thin solid black, corresponding to saturation of Eq.~(\ref{etag})], from the criterion of Ref.~\cite{OneWayPryde} [dashed blue, corresponding to saturation of Eq.~(\ref{howard})], and from the criterion of Ref.~\cite{NhaSciRep} [dotted red, corresponding to saturation of Eq.~(\ref{NhaCri})]. We see that our simple criterion based on type-i pseudospin measurements is quite powerful in revealing $A \rightarrow B$ steerability down to $\eta \geq 1/3$ for small initial squeezing $s$, and is in particular better than the criterion of Eq.~(\ref{howard}) based on the same measurements for larger $s$. We also identify a region (shaded yellow in Fig.~\ref{lossa}) where our analysis certifies steerability not previously detected by any other criterion based on either Gaussian or non-Gaussian (two-outcome) measurements. For $s \gtrsim 0.9$, however, conventional Gaussian measurements are more suited for steering detection than non-Gaussian ones in the considered state. On the other hand, application of our criterion (\ref{mom}) with type-ii pseudospin measurements is ineffective, as it identifies an $A \rightarrow B$ steerable region which is in fact smaller than the one identified by Gaussian measurements, Eq.~(\ref{etag}).

\subsubsection{General two-mode squeezed thermal states}

\begin{figure*}[t]
\begin{center}
\includegraphics[width=8cm]{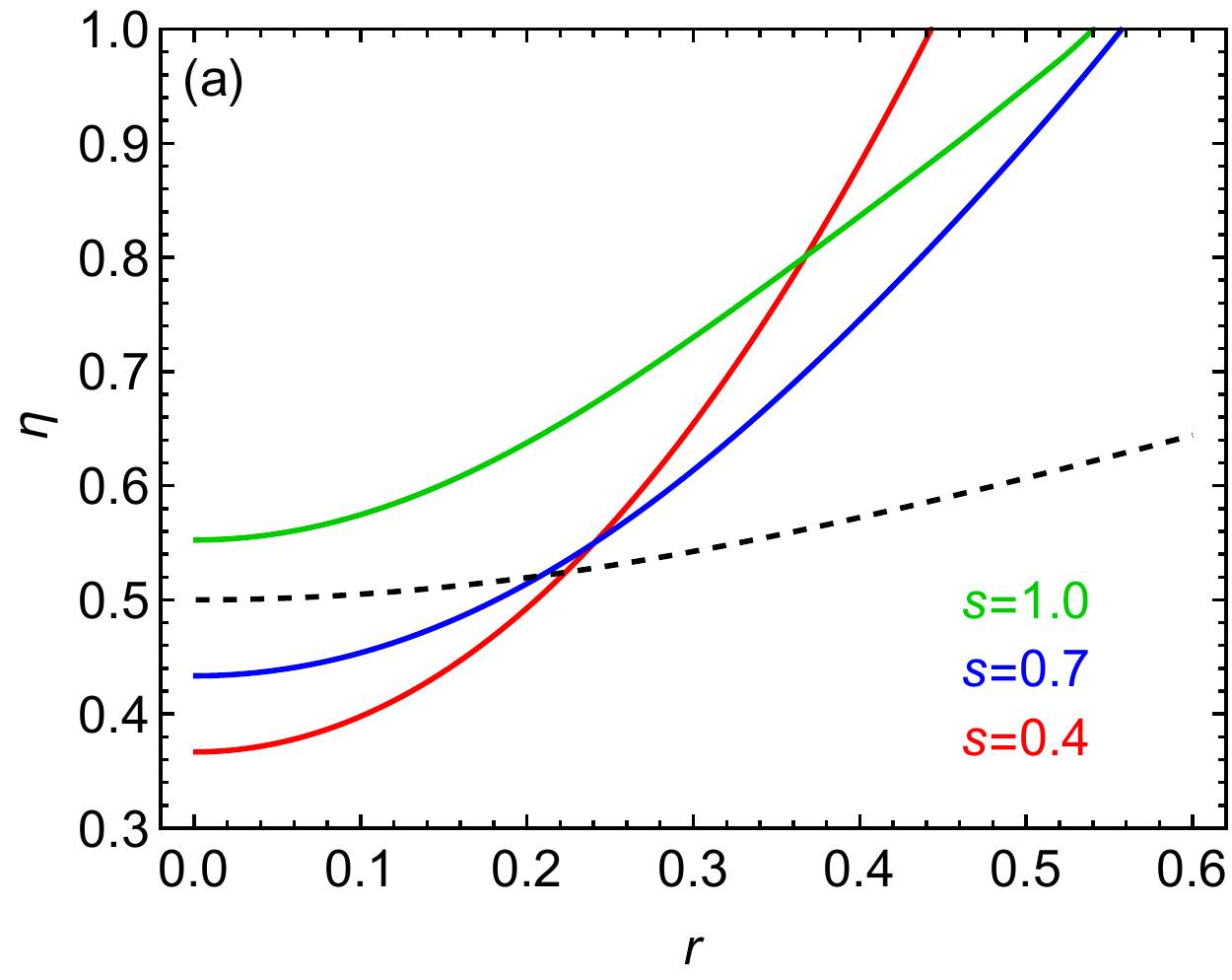} \hspace{1cm}
\includegraphics[width=8cm]{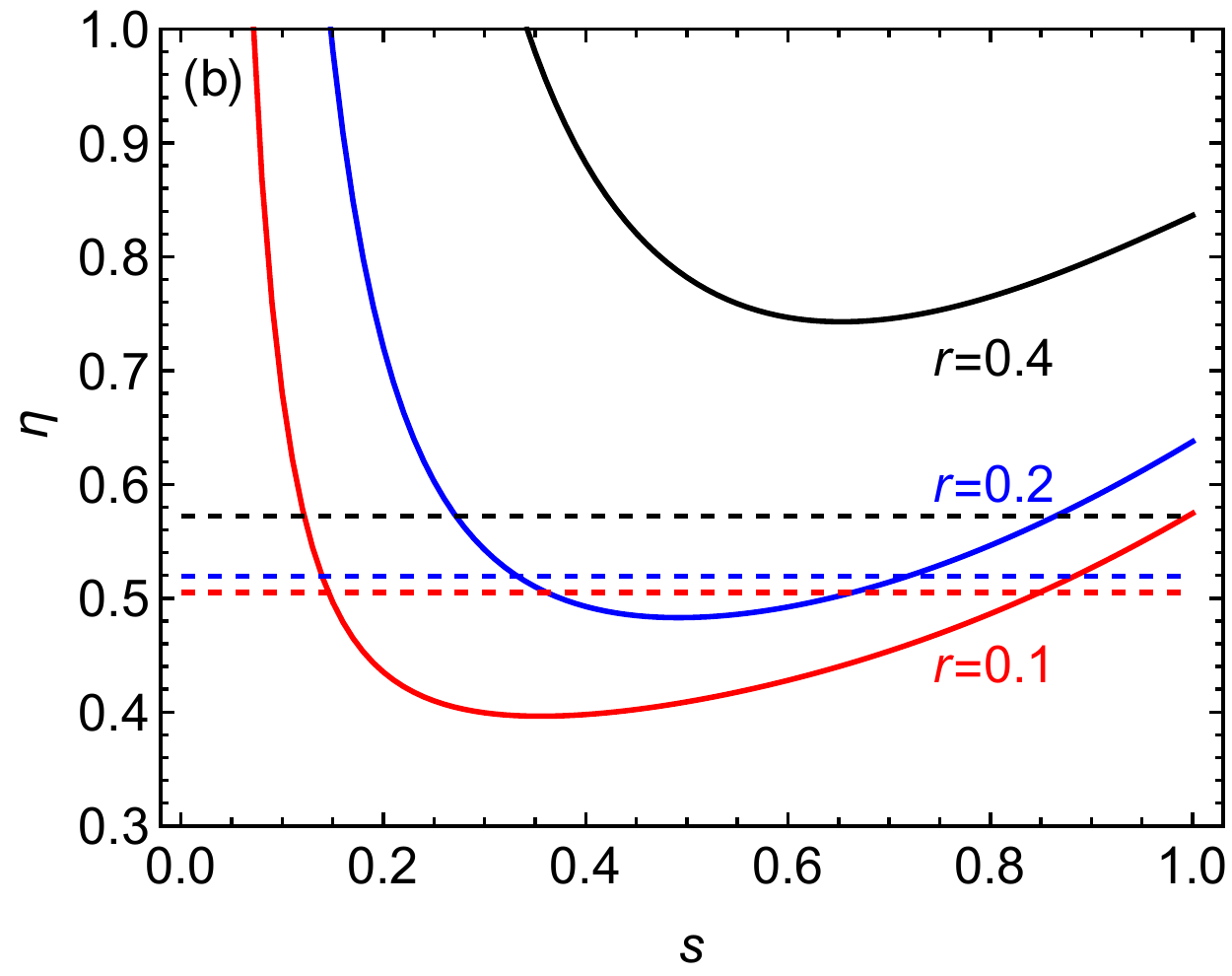}
\protect\caption{(Color online) EPR steering of TMST states $\hat{\rho}^{\text{TMST}}_{AB}(s,\eta,r)$. (a) Relation between the threshold for the transmissivity $\eta$ to detect steering as a function of the amplifier squeezing $r$, for different initial squeezing $s$. (b) Relation between the threshold for $\eta$ to detect steering as a function of  $s$, for different $r$. Steering is detected in the regions above the corresponding curves.  In both panels, solid lines indicate thresholds for the moment criterion Eq.~(\ref{mom}) using type-i pseudospin measurements, while dashed lines indicate thresholds for the criterion Eq.~(\ref{wise2}) using Gaussian measurements. All the quantities plotted are dimensionless.
\label{nonG} }
\end{center}
\end{figure*}

Having verified that the EPR steering criterion Eq.~(\ref{mom}) with (type-i) pseudospin observables is effective to detect steerability of special Gaussian states beyond the capabilities of Gaussian measurements, we can now extend our analysis to general TMST states, for which no previous steering study based on non-Gaussian measurements has been reported to date. Namely, we consider in general the state $\hat{\rho}^{\text{TMST}}_{AB}(s,\eta,r)$ constructed as in Fig.~\ref{trick}, and investigate EPR steering as a function of the three parameters $s$ (initial squeezing of the EPR source), $\eta$ (transmissivity of the attenuator channel) and $r$ (squeezing parameter of the amplifier channel). Our analysis is based on numerical evaluation of the formulas in Eq.~(\ref{sxyztmst}) for the moment criterion $M_{\hat{\rho}^{\text{TMST}}_{AB}}^{(\text{i})}(s,\eta,r) > 1$ using type-i pseudospin measurements, and comparison with the analytical prescription of Eq.~(\ref{wise2}) relying on Gaussian measurements.

The results are reported in Fig.~\ref{nonG}. Panel (a) shows the dependence of the steering thresholds corresponding to non-Gaussian versus Gaussian measurements as a function of the noise parameters $r$ and $\eta$, for different fixed values of the initial squeezing $s$. Panel (a) shows that, even in the presence of both attenuator and amplifier noises on $A$, EPR steering from Alice to Bob can be demonstrated at lower values of $\eta$ by using type-i pseudospin measurements as opposed to Gaussian ones, in particular in the region of moderate $r$. Panel (b) shows in more detail how the analysis of Fig.~\ref{lossa} gets modified by the presence of the additional amplifier noise induced by ${\cal L}_r$. While the Gaussian thresholds for steerability remain independent of $s$, for any fixed $r$, additional regions of steerability identified by type-i pseudospin measurements appear at intermediate values of $s$.  In general, these results give a quite comprehensive picture of the potential enhancements to EPR steering characterization for Gaussian states due to non-Gaussian measurements, and go significantly beyond specific examples considered in previous literature \cite{OneWayPryde,NhaSciRep}.

\subsubsection{Arbitrary two-mode Gaussian states}

Finally, it would be desirable to extend the previous study to arbitrary two-mode Gaussian states, with covariance matrix $\V_{AB}$ specified by all four independent standard form parameters $a,b,c,d$  as in Eq.~(\ref{gamma}). However, the construction of Sec.~\ref{sec:fock} to obtain the Fock basis elements of a Gaussian state $\hat{\rho}_{AB}$ is special to the TMST case, $d=-c$, and its possible extension beyond this case appears quite nontrivial. The only possibility we have, based on the tools employed in this paper, is to use type-ii pseudospin measurements to investigate the EPR steering of arbitrary two-mode Gaussian states using the moment criterion Eq.~(\ref{mom}), thanks to the explicit expressions of Eq.~(\ref{pxyzabcd}). Unfortunately, as anticipated by the special cases investigated in the previous subsections, it turns out that type-ii pseudospin measurements used in the moment criterion of Eq.~(\ref{mom}) are always inferior to Gaussian measurements used in the variance criterion of Eq.~(\ref{wise2}), for all two-mode Gaussian states. This can be proven by maximizing the quantity $M_{\hat{\rho}_{AB}}^{(\text{ii})}$, Eq.~(\ref{mammt}) entering the left-hand side of the moment criterion (\ref{mom}), under the condition that (\ref{wise2}) is violated, that is, that the state is unsteerable by Gaussian measurements. We find
\begin{equation}
\label{nogobx}
\max_{\{a,b,c,d\}} \left. M_{\hat{\rho}_{AB}}^{(\text{ii})}(a,b,c,d) \right\vert_{a^2 \leq (ab-c^2)(ab-d^2)} = 1\,,
\end{equation}
which is obtained for $a=b=1$, $c=d=0$, i.e., when ${\rho_{AB}}$ reduces to the product of vacuum states for $A$ and $B$, with $\V_{AB}=\openone_A \oplus \openone_B$. This shows that the moment criterion (\ref{mom}) can never detect EPR steering using type-ii pseudospin measurements if the state is not already steerable by Gaussian measurements. In fact, the steerability region as detected by $M_{\hat{\rho}_{AB}}^{(\text{ii})}>1$ is strictly smaller than the one defined by Eq.~(\ref{wise2}), as demonstrated in the instance of Fig.~\ref{lossa}. However, this does not exclude that the type-ii pseudospin operators might be useful to detect EPR steering of Gaussian states beyond Gaussian measurements if other criteria, possibly involving higher order moments, are considered.

\section{Steering of continuous variable non-Gaussian Werner states}\label{sec:werner}

Up to now, we focused on the investigation of EPR steering for Gaussian states using non-Gaussian measurements.
A next logical step is to include also non-Gaussian states into consideration. Here, we probe this scenario by
analyzing steerability for a paradigmatic example of mixed non-Gaussian states given by the class of CV
Werner states \cite{Mista_02b}. In past literature, the CV Werner states have been studied from the point of view of
inseparability, nonlocality and optical nonclassicality \cite{Mista_02b}, as well as quantum discord \cite{Tatham_12}.
In this Section we complement the list by analyzing EPR steering of these states as detected by the inequality (\ref{mom}) with type-i pseudospin operators
(\ref{sxyz}).

A CV Werner state is defined as the convex mixture
\begin{equation}\label{rhoW}
\hat{\rho}^{W}=p\,\hat{\rho}^{\text{EPR}}_{AB}(s)+(1-p)\,[\hat{\rho}^{\rm th}_{A}(u)\otimes\hat{\rho}^{\rm th}_{B}(u)],
\end{equation}
where $0\leq p\leq 1$, $\hat{\rho}^{\text{EPR}}_{AB}(s)$ is the two-mode squeezed vacuum state (\ref{rhoTMSV}),
and
\begin{equation}\label{rhoth}
\hat{\rho}^{\rm th}_{j}(u)=[1-\tanh^{2}(u)]\sum_{m=0}^{\infty}
\tanh^{2m}(u)|m\rangle_{j}\langle m|,\quad j=A,B
\end{equation}
is a thermal state with $\tanh^{2}(u)=\langle n_{j}\rangle/(1+\langle n_{j}\rangle)$, where
$\langle n_{j}\rangle$ is the mean number of thermal photons in mode $j$. For $u=s$, the state
(\ref{rhoW}) can be interpreted as originating from transmission of one mode of the two-mode squeezed
vacuum state (\ref{rhoTMSV}) through a non-Gaussian channel which, with probability $p$, transmits the mode unaltered
and, with probability $1-p$, replaces the mode with a thermal state (\ref{rhoth}) with $u=s$. In addition,
in the limit  $s\rightarrow \infty$ the latter Werner state provides a direct analogy to the original
discrete-variable Werner state \cite{Werner_89}, because it becomes a mixture of a maximally entangled EPR state and a maximally mixed
state in the infinite-dimensional Hilbert space.

\begin{figure}[t]
\begin{center}
\includegraphics[width=7.5cm]{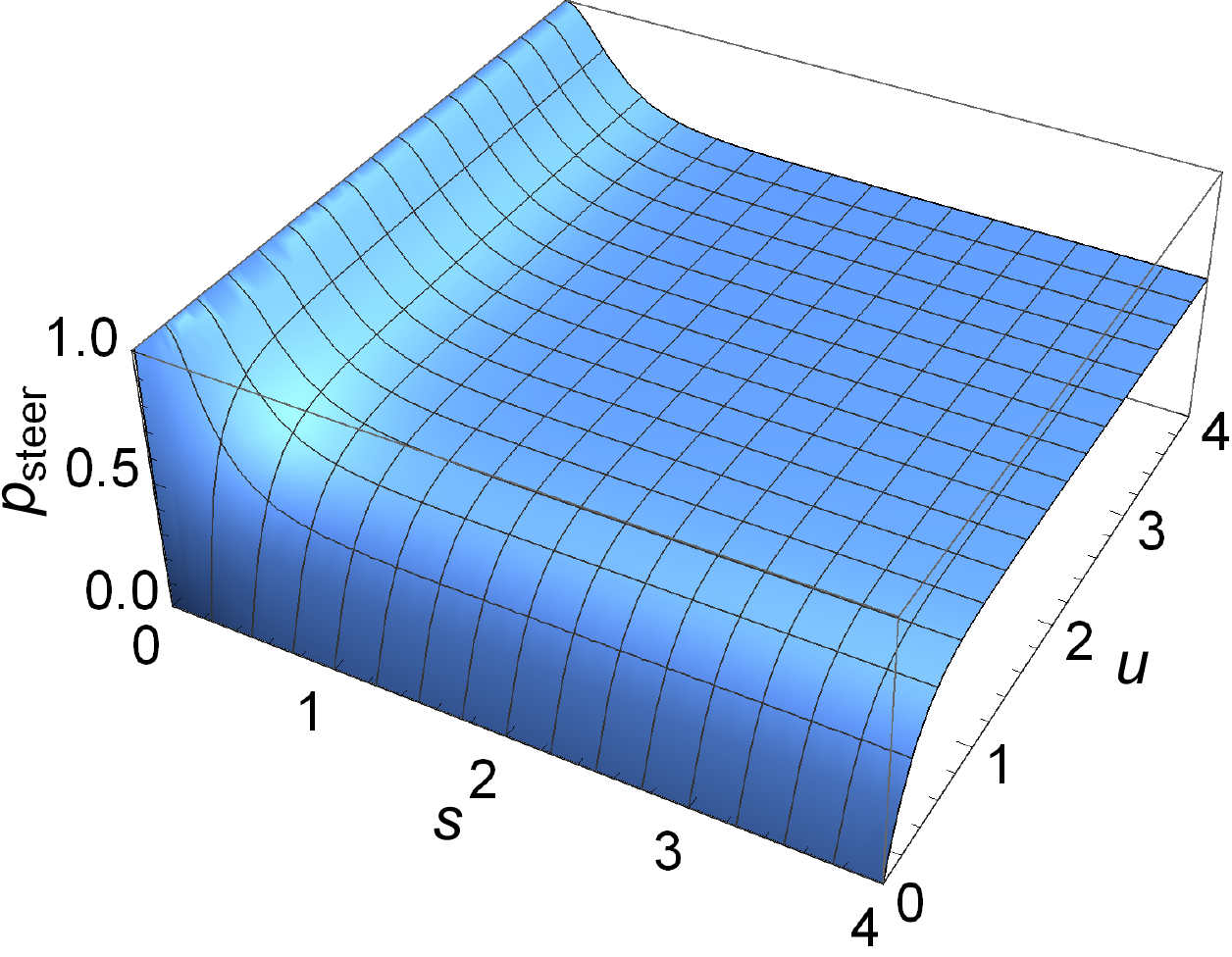}
\protect\caption{(Color online) Threshold probability $p_{\rm steer} \equiv p_{\rm steer}^{(\text{i})}$ [Eq.~(\ref{psteer})] characterizing the EPR steering of the CV Werner state (\ref{rhoW}) as a
function of the squeezing parameter $s$ and the thermal noise parameter $u$. According to inequality (\ref{mom}), the state is steerable by type-i pseudospin measurements if $p>p_{\rm steer}^{(\text{i})}$.
All the quantities plotted are dimensionless.
\label{figpsteer}}
\end{center}
\end{figure}

Moving to the determination of the region of parameters $p$,$s$, and $u$, for which the steering inequality (\ref{mom}) is satisfied,
we first need to derive the expectation values $\langle \hat{S}^j_A \otimes \hat{S}^j_B\rangle$, $j=x,y,z$, of pairs of type-i pseudospin operators (\ref{sxyz}) on the
CV Werner state (\ref{rhoW}). Straightforward algebra reveals that the expectation values attain
the following simple form \cite{Mista_02b},
\begin{equation}\label{sxyzW}
\begin{split}
\langle \hat{S}^x_A \otimes \hat{S}^x_B \rangle&=-\langle \hat{S}^y_A \otimes \hat{S}^y_B \rangle=p v\,,  \\
\langle \hat{S}^z_A \otimes \hat{S}^z_B \rangle&=pw+1-w\,,
\end{split}
\end{equation}
where $v=\tanh(2s)$ and $w=\tanh^{2}(2u)$. Making use of the expectation values (\ref{sxyzW}), one then finds that the steering inequality (\ref{mom}) boils down to
\begin{equation}\label{momW}
p^2+\frac{2w(1-w)}{2v^2+w^2}p-\frac{w(2-w)}{2v^2+w^2}>0,
\end{equation}
which is equivalent to
\begin{equation}\label{psteer}
p>p_{\rm steer}^{(\text{i})}\equiv\frac{\sqrt{w(w-2v^2w+4v^2)}-w(1-w)}{2v^2+w^2}.
\end{equation}
The region of fulfilment of the steering inequality (\ref{mom}) for the CV Werner state (\ref{rhoW}) with type-i pseudospin measurements is
depicted in Fig.~\ref{figpsteer}. By comparing it with the results of \cite{Mista_02}, one can see that the threshold $p_{\rm steer}^{(\text{i})}$ for detecting steering is lower than the one for detecting Bell nonlocality, as it should be expected given the hierarchy existing between these two forms of nonclassical correlations. Furthermore, for $u=s$ and in the strong squeezing limit, the inequality (\ref{psteer})
reduces to $p>1/\sqrt{3}$, which coincides with the threshold for steering of a two-qubit Werner state when
Alice has exactly three inputs \cite{Cavalcanti_09}.

We can also check the steerable region predicted by the inequality (\ref{mom}) when using type-ii pseudospin operators (\ref{pxyz}). By linearity, the expectation values
 $\langle \hat{\Pi}^j_A \otimes \hat{\Pi}^j_B\rangle$, $j=x,y,z$, of pairs of type-ii pseudospin operators on the CV Werner state  can be obtained as convex combinations of the corresponding expectation values on the two Gaussian states entering the definition (\ref{rhoW}). Exploiting Eqs.~(\ref{pxyzabcd}), we find
\begin{equation}\label{pxyzW}
\begin{split}
\langle \hat{\Pi}^x_A \otimes \hat{\Pi}^x_B \rangle&=\langle \hat{\Pi}^y_A \otimes \hat{\Pi}^y_B \rangle=\frac{2p}{\pi} \text{gd}(2s),  \\
\langle \hat{\Pi}^z_A \otimes \hat{\Pi}^z_B \rangle&=p+(1-p)\text{sech}^2(2u)\,.
\end{split}
\end{equation}
Using the expectation values (\ref{pxyzW}), one then finds that the steering inequality (\ref{mom}) is fulfilled using type-ii pseudospin measurements when
\begin{equation}\label{psteer2}
p>p_{\rm steer}^{(\text{ii})}\equiv \frac{\sqrt{w \left(\pi ^2 w+8 (2-w) \text{gd}^2(2 s)\right)}-\pi  w(1-w)}{(8/\pi) \text{gd}^2(2 s)+\pi  w^2}.
\end{equation}
We find that the threshold $p_{\rm steer}^{(\text{ii})}$ is  only slightly higher than $p_{\rm steer}^{(\text{i})}$ in the regime of small $s$, however they both converge to $1/\sqrt{3}$ in the asymptotic limit $u=s \rightarrow \infty$. Therefore both types of pseudospin observables are equally effective in this relevant regime.

\begin{figure}[t]
\begin{center}
\includegraphics[width=8cm]{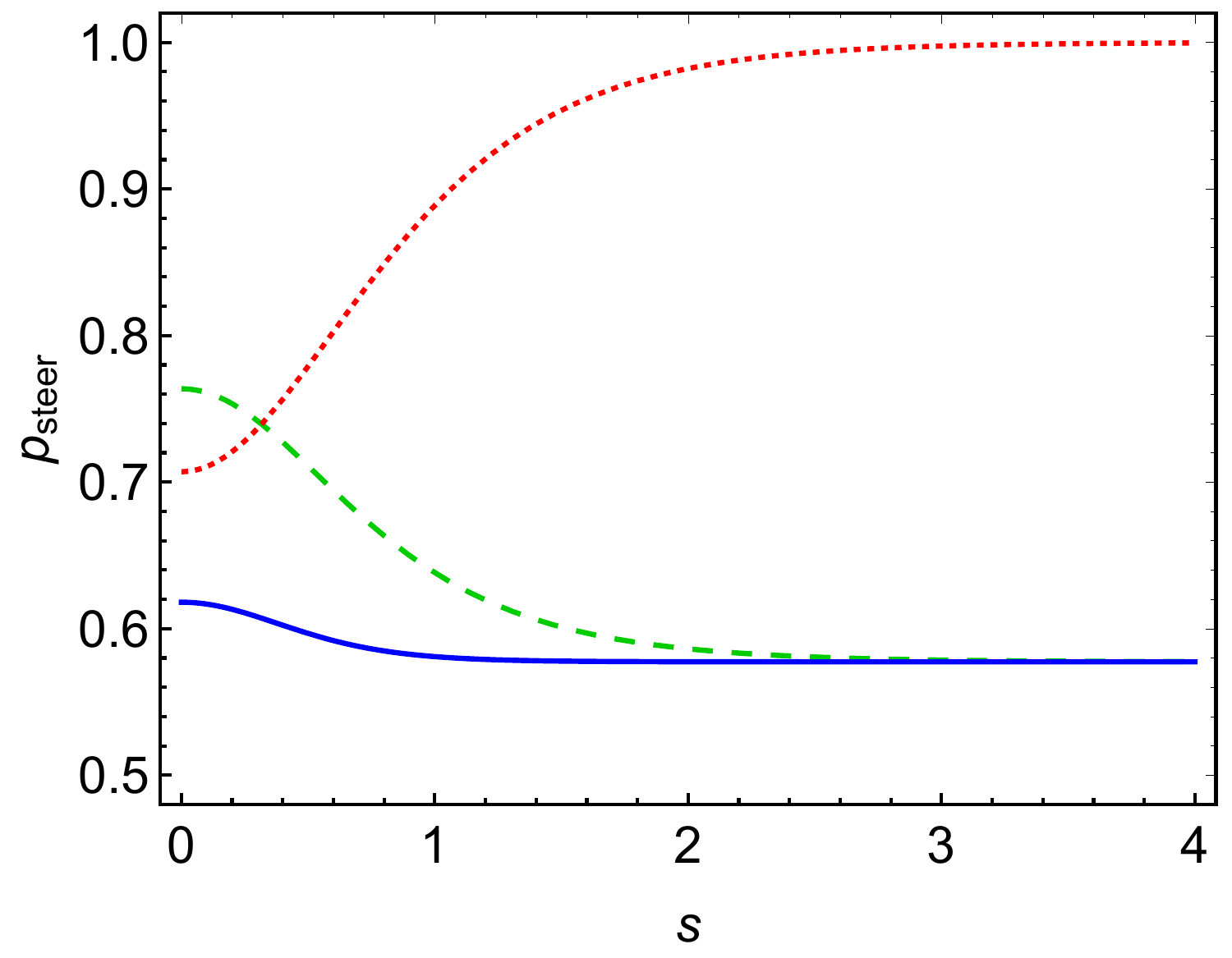}
\protect\caption{(Color online) Threshold probabilities characterizing the EPR steering of the CV Werner state (\ref{rhoW}) with $u=s$. The state is steerable by type-i pseudospin measurements when $p>p_{\rm steer}^{(\text{i})}$  [Eq.~(\ref{psteer})] (solid blue line), by type-ii pseudospin measurements when $p>p_{\rm steer}^{(\text{ii})}$ [Eq.~(\ref{psteer2})] (dashed green line), and by Gaussian measurements when $p>p_{\rm steer}^{(\text{G})}$ (dotted red line) [Eq.~(\ref{Gpsteer})]. All the quantities plotted are dimensionless.
\label{figWus}}
\end{center}
\end{figure}

For the sake of comparison, we can finally look at the region of parameters $p$, $s$, and $u$ in which the CV Werner state is steerable by Gaussian measurements. For this purpose, we need the covariance matrix of the state $\hat{\rho}^{W}$, which is simply given by the linear combination of the covariance matrices of the Gaussian states appearing in the convex mixture (\ref{rhoW}),
\begin{equation}\label{VW}
\V^W_{AB}= p \V_{AB}^{\text{EPR}}(s) + (1-p) \cosh(2u) (\openone_{A}\oplus \openone_{B})\,.
\end{equation}
Explicitly, $\V^W_{AB}$ is in the standard form (\ref{gamma}) with $a=b=p \cosh(2s) + (1-p) \cosh(2u)$ and $c=-d=p \sinh(2s)$. According to (\ref{wise2}), the state is steerable by Gaussian measurements when $a>a^2-c^2$, which amounts to
\begin{equation}\label{Gpsteer}
p>p_{\rm steer}^{(\text{G})}\equiv \mbox{$\frac{\sqrt{c_s^2 \left(1-2 c_u\right)^2-2 c_s c_u+c_u \left(4-3 c_u\right)}+c_s \left(2 c_u-1\right)-2 c_u^2+c_u}{4 c_s c_u-2 \left(c_u^2+1\right)}$}\,,
\end{equation}
where $c_s=\cosh(2s)$ and $c_u=\cosh(2u)$.
Remarkably, one sees that $p_{\rm steer}^{(\text{G})} > p_{\rm steer}^{(\text{i})}$ for any $s,u>0$, meaning that non-Gaussian pseudospin measurements are always superior to Gaussian measurements for the characterization of EPR steering in the non-Gaussian CV Werner states. In particular, when $u=s$ we find $p_{\rm steer}^{(\text{G})}=1/\sqrt{1+\text{sech}(2s)}$, which tends to $1$ in the limit $s \rightarrow \infty$, meaning that --- although the state is steerable for $p>1/\sqrt{3}$ as confirmed by either (\ref{psteer}) or (\ref{psteer2}) --- Gaussian measurements can never detect steering in this asymptotic case unless $p=1$, i.e.~when the state (\ref{rhoW}) trivially reduces to the EPR state (\ref{rhoTMSV}). A comparison between the EPR steering thresholds (\ref{psteer}), (\ref{psteer2}), and (\ref{Gpsteer}) for the CV Werner state with $u=s$ is provided in Fig.~\ref{figWus}.

\section{Conclusions}\label{sec:concl}

In this paper we investigated EPR steering \cite{wiseman} of continuous variable bipartite states, as revealed by a simple nonlinear criterion \cite{kogias3} involving the second moments of pseudospin measurements \cite{chen,gour1}. Our analysis led to the identification of sizeable regions of parameters in which Gaussian states, in particular two-mode squeezed thermal states, can only be steered by non-Gaussian measurements, complementing and extending recent findings \cite{OneWayPryde,NhaSciRep}. We also showed that non-Gaussian (pseudospin) measurements are more effective than Gaussian (quadrature) measurements for witnessing steering of non-Gaussian continuous variable Werner states \cite{Mista_02b}, whose steerability properties were found comparable to their discrete-variable counterparts \cite{Cavalcanti_09}. While pseudospin observables are experimentally hard to measure with current technology, our results can stimulate further research to identify accessible non-Gaussian measurements for enhanced steering detection in Gaussian and non-Gaussian states. Since steering is a fundamental resource for quantum communication \cite{resource,branciard,walk,Reid13,He15,He2015,kogias,kogiasqss}, this can lead in turn to further advances in the engineering of secure quantum network architectures based on continuous variable systems. It would be interesting in the future to investigate generalizations of our study to multipartite settings \cite{he,seiji2015,cavalcanti-nc}, analyzing in particular relaxations to strict monogamy inequalities for steering which hold specifically in the all-Gaussian setting \cite{reid13a,kogiasmono,Lami2016,Deng17}.

In the first part of the paper, we also obtained a result of independent interest, that is the Fock representation of arbitrary two-mode squeezed thermal states.
In this respect, recall that a standard approach
\cite{Dodonov_84,Dodonov_94,Fiurasek_review01} to the derivation of the elements of a
quantum state in the Fock basis makes use of the fact that the Husimi $Q$ quasi-probability distribution of the
state is proportional to a generating function of the elements. For a Gaussian state, the generating function
is Gaussian and thus it is at the same time a generating function for multidimensional
Hermite polynomials \cite{Bateman_53}. This implies that, up to a normalization factor, the Fock basis
elements of Gaussian states are equal to multidimensional Hermite polynomials (see \hyperref[sec:app]{Appendix} for details).
Since for higher orders the polynomials are obtained as multiple derivatives of a multivariate Gaussian function, they are very complex, and thus in practical
tasks one has to evaluate them numerically using a recurrence relation \cite{Mista_14}. Here, we undertook a different route
by calculating the Fock basis elements directly with the help of an expression of a two-mode squeezed thermal state via a two-mode squeezed
vacuum state with one mode exposed to a phase-insensitive Gaussian channel, and a decomposition of the latter channel
into a sequence of a pure-loss channel and a quantum-limited amplifier \cite{Caruso_06,Garcia-Patron_12,Pirandola_14}. This led to a rather simple formula
for the density matrix elements in terms of a single finite sum as given in Eq.~(\ref{Fockfinal}).
On a more general level, given the correspondence outlined above, our approach may also serve as an inspiration to derive new relations for multidimensional
Hermite polynomials. An explicit instance is discussed in the \hyperref[sec:app]{Appendix}.

\acknowledgments{Discussions with C.~Budroni, G.~Gour, O.~G\"uhne, I.~Kogias, S.-W.~Ji, H.~Nha, Z.-Y.~Ou, X.-L.~Su, and V.~Vennin are gratefully acknowledged. Y.X. and Q.H. acknowledge financial  support from the Ministry of Science and Technology of China (Grant No. 2016YFA0301302) and the National Natural Science Foundation of China (Grants No. 11622428, No. 61475006, and No. 61675007). B.X, T.T., and G.A. acknowledge financial support from the European Research Council (ERC) under the Starting Grant GQCOP (Grant No.~637352) and the Foundational Questions Institute (fqxi.org) under the Physics of the Observer Programme (Grant No.~FQXi-RFP-1601).}

\appendix*
\setcounter{equation}{0}
\section{Compact expression for multidimensional \\ Hermite polynomials}\label{sec:app}

The method used in Sec.~\ref{sec:fock} for the derivation of the Fock basis elements of TMST states, given by Eq.~(\ref{Fockfinal}),
can be a cornerstone for a wider algebra programme aimed at the derivation of new compact expressions for
multidimensional Hermite polynomials based on quantum mechanics.

Making use of the results of Refs.~\cite{Dodonov_84,Dodonov_94,Fiurasek_review01}, it can be
shown \cite{Mista_14} that, for a TMST state $\hat{\rho}^{\text{TMST}}_{AB}$ with covariance matrix $\V_{AB}$ given by the right-hand
side of Eq.~(\ref{gamma}), where $d=-c$, the Fock basis elements can be written formally as %
\begin{eqnarray}\label{FockHermite}
\langle m_{1}m_{2}|\hat{\rho}^{\text{TMST}}_{AB}|n_{1}n_{2}\rangle=
\frac{4H_{m_1,m_2,n_1,n_2}^{(\gr\Theta)}(\gr 0)}{\sqrt{\text{det}\left(\V_{AB}+\openone\right)}\sqrt{m_1!m_2!n_1!n_2!}}.\nonumber\\
\end{eqnarray}
Here $H_{m_1,m_2,n_1,n_2}^{(\gr\Theta)}(\gr 0)$ is the four-dimensional Hermite polynomial at the
origin, which can be calculated from the expression \cite{Bateman_53}
\begin{equation}
\begin{split}
H_{m_1,m_2,n_1,n_2}^{(\gr\Theta)}(\gr x)&=(-1)^{\sum_{i=1}^{2}
n_i+m_i}\exp\left(\frac{1}{2}\gr x\T \gr \Theta \gr x\right)  \\
&\times \frac{\partial^{\sum_{i=1}^2
n_i+m_i}}{\partial {x_1^{m_1}}\partial {x_2^{m_2}}\partial
{{x}_3^{n_1}}\partial
{{x}_4^{n_2}}}\exp\left(-\frac{1}{2}\gr x\T \gr \Theta \gr x\right), \\
\end{split}
\end{equation}
and $\gr \Theta$ is a real symmetric matrix defining the polynomial, which is of the form:
\begin{equation}\label{R}
\gr \Theta=-\left(\begin{array}{cccc}
0 & e & f & 0\\
e & 0 & 0 & g\\
f & 0 & 0 & e\\
0 & g & e & 0\end{array}\right),
\end{equation}
where
\begin{equation*}\label{uvw}
\begin{split}
&e=\frac{2c}{(a+1)(b+1)-c^2}\,,\nonumber \\ &f=\frac{(a-1)(b+1)-c^2}{(a+1)(b+1)-c^2}\,, \quad
g=\frac{(a+1)(b-1)-c^2}{(a+1)(b+1)-c^2}\,.\nonumber \end{split}
\end{equation*}

By comparing the right-hand sides of Eq.~(\ref{FockHermite}) and Eq.~(\ref{Fockfinal}) and taking into account the
relation
\begin{eqnarray}
\frac{1-\varsigma^{2}}{\cosh^{2}(r)}&=&\frac{4}{\sqrt{\text{det}\left(\V_{AB}+\openone\right)}},\label{det}
\end{eqnarray}
which follows from Eqs.~(\ref{ac}), one finds that the Hermite polynomials $H_{m_1,m_2,n_1,n_2}^{(\gr\Theta)}(\gr 0)$ can be expressed as
\begin{eqnarray}\label{Hermitenew}
&&\!\!H_{m_1,m_2,n_1,n_2}^{(\gr\Theta)}(\gr 0)\nonumber\\
&&=\sqrt{m_1!m_2!n_1!n_2!}\delta_{m_1+n_2,n_{1}+m_{2}}\left[\frac{\varsigma \sqrt{\eta}}{\cosh(r)}\right]^{m_{2}+n_{2}}\!\!\!\left[\tanh(r)\right]^{2(m_{1}-m_{2})} \nonumber \\ &&\ \ \ \times
\sum_{k=\mathrm{max}\{0,m_{2}-m_{1}\}}^{\mathrm{min}\{m_{2},n_{2}\}}\sqrt{{m_{2}\choose k}{n_{2}\choose k}{m_{1}\choose m_{2}-k}{n_{1}\choose n_{2}-k}}\nonumber\\
&&\ \ \ \times\left[\sqrt{\frac{1-\eta}{\eta}}\sinh(r)\right]^{2k}.
\end{eqnarray}
If we now reverse Eqs.~(\ref{ac}), we can express after some algebra the parameters $\sqrt{\eta}, \varsigma$, as well as all functions
of the squeezing parameter $r$ which appear on the right-hand side of Eq.~(\ref{Hermitenew}), as functions of
the covariance matrix parameters $a,b$ and $c$, which finally yields the following formula for four-dimensional Hermite polynomials defined by the matrix $\gr \Theta$ in Eq.~(\ref{R}):
\begin{eqnarray}\label{Hermitenewfinal}
&&\!\!\!\!\!\!\!\!H_{m_1,m_2,n_1,n_2}^{(\gr\Theta)}(\gr 0)=\sqrt{m_1!m_2!n_1!n_2!}\delta_{m_1+n_2,n_{1}+m_{2}}e^{m_{2}+n_{2}}f^{m_{1}-m_{2}} \nonumber \\ &&\ \ \ \times
\sum_{k=\mathrm{max}\{0,m_{2}-m_{1}\}}^{\mathrm{min}\{m_{2},n_{2}\}}\sqrt{{m_{2}\choose k}{n_{2}\choose k}{m_{1}\choose m_{2}-k}{n_{1}\choose n_{2}-k}}\left(\frac{fg}{e^2}\right)^{k}.\nonumber\\
\end{eqnarray}

Note that, in the usual practice, Hermite polynomials are evaluated numerically using a recurrence relation.
Our last formula (\ref{Hermitenewfinal}) shows that, in some cases, this is not necessary and one
can obtain them just as a single finite sum instead. This can make analytical calculations with multidimensional
Hermite polynomials more tractable and numerical calculations more efficient.

%


\begin{thebibliography}{69}%
\makeatletter
\providecommand \@ifxundefined [1]{%
 \@ifx{#1\undefined}
}%
\providecommand \@ifnum [1]{%
 \ifnum #1\expandafter \@firstoftwo
 \else \expandafter \@secondoftwo
 \fi
}%
\providecommand \@ifx [1]{%
 \ifx #1\expandafter \@firstoftwo
 \else \expandafter \@secondoftwo
 \fi
}%
\providecommand \natexlab [1]{#1}%
\providecommand \enquote  [1]{``#1''}%
\providecommand \bibnamefont  [1]{#1}%
\providecommand \bibfnamefont [1]{#1}%
\providecommand \citenamefont [1]{#1}%
\providecommand \href@noop [0]{\@secondoftwo}%
\providecommand \href [0]{\begingroup \@sanitize@url \@href}%
\providecommand \@href[1]{\@@startlink{#1}\@@href}%
\providecommand \@@href[1]{\endgroup#1\@@endlink}%
\providecommand \@sanitize@url [0]{\catcode `\\12\catcode `\$12\catcode
  `\&12\catcode `\#12\catcode `\^12\catcode `\_12\catcode `\%12\relax}%
\providecommand \@@startlink[1]{}%
\providecommand \@@endlink[0]{}%
\providecommand \url  [0]{\begingroup\@sanitize@url \@url }%
\providecommand \@url [1]{\endgroup\@href {#1}{\urlprefix }}%
\providecommand \urlprefix  [0]{URL }%
\providecommand \Eprint [0]{\href }%
\providecommand \doibase [0]{http://dx.doi.org/}%
\providecommand \selectlanguage [0]{\@gobble}%
\providecommand \bibinfo  [0]{\@secondoftwo}%
\providecommand \bibfield  [0]{\@secondoftwo}%
\providecommand \translation [1]{[#1]}%
\providecommand \BibitemOpen [0]{}%
\providecommand \bibitemStop [0]{}%
\providecommand \bibitemNoStop [0]{.\EOS\space}%
\providecommand \EOS [0]{\spacefactor3000\relax}%
\providecommand \BibitemShut  [1]{\csname bibitem#1\endcsname}%
\let\auto@bib@innerbib\@empty
\bibitem [{\citenamefont {Dowling}\ and\ \citenamefont
  {Milburn}(2003)}]{Dowling}%
  \BibitemOpen
  \bibfield  {author} {\bibinfo {author} {\bibfnamefont {J.~P.}\ \bibnamefont
  {Dowling}}\ and\ \bibinfo {author} {\bibfnamefont {G.~J.}\ \bibnamefont
  {Milburn}},\ }\href {\doibase10.1098/rsta.2003.1227} {\bibfield  {journal}
  {\bibinfo  {journal} {Phil. Trans. Roy. Soc. A}\ }\textbf {\bibinfo {volume}
  {361}},\ \bibinfo {pages} {1655} (\bibinfo {year} {2003})}\BibitemShut
  {NoStop}%
\bibitem [{\citenamefont {Streltsov}\ \emph {et~al.}(2017)\citenamefont
  {Streltsov}, \citenamefont {Adesso},\ and\ \citenamefont
  {Plenio}}]{streltsov2016quantum}%
  \BibitemOpen
  \bibfield  {author} {\bibinfo {author} {\bibfnamefont {A.}~\bibnamefont
  {Streltsov}}, \bibinfo {author} {\bibfnamefont {G.}~\bibnamefont {Adesso}}, \
  and\ \bibinfo {author} {\bibfnamefont {M.~B.}\ \bibnamefont {Plenio}},\
  }\href@noop {} {\bibfield  {journal} {\bibinfo  {journal}
  {arXiv:1609.02439v3}\ } (\bibinfo {year} {2017})}\BibitemShut {NoStop}%
\bibitem [{\citenamefont {Horodecki}\ \emph {et~al.}(2009)\citenamefont
  {Horodecki}, \citenamefont {Horodecki}, \citenamefont {Horodecki},\ and\
  \citenamefont {Horodecki}}]{ent}%
  \BibitemOpen
  \bibfield  {author} {\bibinfo {author} {\bibfnamefont {R.}~\bibnamefont
  {Horodecki}}, \bibinfo {author} {\bibfnamefont {P.}~\bibnamefont
  {Horodecki}}, \bibinfo {author} {\bibfnamefont {M.}~\bibnamefont
  {Horodecki}}, \ and\ \bibinfo {author} {\bibfnamefont {K.}~\bibnamefont
  {Horodecki}},\ }\href@noop {} {\bibfield  {journal} {\bibinfo  {journal}
  {Rev. Mod. Phys.}\ }\textbf {\bibinfo {volume} {81}},\ \bibinfo {pages} {865}
  (\bibinfo {year} {2009})}\BibitemShut {NoStop}%
\bibitem [{\citenamefont {Modi}\ \emph {et~al.}(2012)\citenamefont {Modi},
  \citenamefont {Brodutch}, \citenamefont {Cable}, \citenamefont {Paterek},\
  and\ \citenamefont {Vedral}}]{disc}%
  \BibitemOpen
  \bibfield  {author} {\bibinfo {author} {\bibfnamefont {K.}~\bibnamefont
  {Modi}}, \bibinfo {author} {\bibfnamefont {A.}~\bibnamefont {Brodutch}},
  \bibinfo {author} {\bibfnamefont {H.}~\bibnamefont {Cable}}, \bibinfo
  {author} {\bibfnamefont {T.}~\bibnamefont {Paterek}}, \ and\ \bibinfo
  {author} {\bibfnamefont {V.}~\bibnamefont {Vedral}},\ }\href@noop {}
  {\bibfield  {journal} {\bibinfo  {journal} {Rev. Mod. Phys.}\ }\textbf
  {\bibinfo {volume} {84}},\ \bibinfo {pages} {1655} (\bibinfo {year}
  {2012})}\BibitemShut {NoStop}%
\bibitem [{\citenamefont {Adesso}\ \emph {et~al.}(2016)\citenamefont {Adesso},
  \citenamefont {Bromley},\ and\ \citenamefont {Cianciaruso}}]{ABC}%
  \BibitemOpen
  \bibfield  {author} {\bibinfo {author} {\bibfnamefont {G.}~\bibnamefont
  {Adesso}}, \bibinfo {author} {\bibfnamefont {T.~R.}\ \bibnamefont {Bromley}},
  \ and\ \bibinfo {author} {\bibfnamefont {M.}~\bibnamefont {Cianciaruso}},\
  }\href {http://stacks.iop.org/1751-8121/49/i=47/a=473001} {\bibfield
  {journal} {\bibinfo  {journal} {J. Phys. A.: Math. Theor.}\ }\textbf
  {\bibinfo {volume} {49}},\ \bibinfo {pages} {473001} (\bibinfo {year}
  {2016})}\BibitemShut {NoStop}%
\bibitem [{\citenamefont {Reid}\ \emph {et~al.}(2009)\citenamefont {Reid},
  \citenamefont {Drummond}, \citenamefont {Bowen}, \citenamefont {Cavalcanti},
  \citenamefont {Lam}, \citenamefont {Bachor}, \citenamefont {Andersen},\ and\
  \citenamefont {Leuchs}}]{steeringreview1}%
  \BibitemOpen
  \bibfield  {author} {\bibinfo {author} {\bibfnamefont {M.~D.}\ \bibnamefont
  {Reid}}, \bibinfo {author} {\bibfnamefont {P.~D.}\ \bibnamefont {Drummond}},
  \bibinfo {author} {\bibfnamefont {W.~P.}\ \bibnamefont {Bowen}}, \bibinfo
  {author} {\bibfnamefont {E.~G.}\ \bibnamefont {Cavalcanti}}, \bibinfo
  {author} {\bibfnamefont {P.~K.}\ \bibnamefont {Lam}}, \bibinfo {author}
  {\bibfnamefont {H.~A.}\ \bibnamefont {Bachor}}, \bibinfo {author}
  {\bibfnamefont {U.~L.}\ \bibnamefont {Andersen}}, \ and\ \bibinfo {author}
  {\bibfnamefont {G.}~\bibnamefont {Leuchs}},\ }\href@noop {} {\bibfield
  {journal} {\bibinfo  {journal} {Rev. Mod. Phys.}\ }\textbf {\bibinfo {volume}
  {81}},\ \bibinfo {pages} {1727} (\bibinfo {year} {2009})}\BibitemShut
  {NoStop}%
\bibitem [{\citenamefont {Cavalcanti}\ and\ \citenamefont
  {Skrzypczyk}(2017)}]{steeringreview2}%
  \BibitemOpen
  \bibfield  {author} {\bibinfo {author} {\bibfnamefont {D.}~\bibnamefont
  {Cavalcanti}}\ and\ \bibinfo {author} {\bibfnamefont {P.}~\bibnamefont
  {Skrzypczyk}},\ }\href {\doibase10.1088/1361-6633/80/2/024001} {\bibfield
  {journal} {\bibinfo  {journal} {Rep. Prog. Phys.}\ }\textbf {\bibinfo
  {volume} {80}},\ \bibinfo {pages} {024001} (\bibinfo {year}
  {2017})}\BibitemShut {NoStop}%
\bibitem [{\citenamefont {Brunner}\ \emph {et~al.}(2014)\citenamefont
  {Brunner}, \citenamefont {Cavalcanti}, \citenamefont {Pironio}, \citenamefont
  {Scarani},\ and\ \citenamefont {Wehner}}]{nonloc}%
  \BibitemOpen
  \bibfield  {author} {\bibinfo {author} {\bibfnamefont {N.}~\bibnamefont
  {Brunner}}, \bibinfo {author} {\bibfnamefont {D.}~\bibnamefont {Cavalcanti}},
  \bibinfo {author} {\bibfnamefont {S.}~\bibnamefont {Pironio}}, \bibinfo
  {author} {\bibfnamefont {V.}~\bibnamefont {Scarani}}, \ and\ \bibinfo
  {author} {\bibfnamefont {S.}~\bibnamefont {Wehner}},\ }\href@noop {}
  {\bibfield  {journal} {\bibinfo  {journal} {Rev. Mod. Phys.}\ }\textbf
  {\bibinfo {volume} {86}},\ \bibinfo {pages} {419} (\bibinfo {year}
  {2014})}\BibitemShut {NoStop}%
\bibitem [{\citenamefont {Schr\"odinger}(1935)}]{schr}%
  \BibitemOpen
  \bibfield  {author} {\bibinfo {author} {\bibfnamefont {E.}~\bibnamefont
  {Schr\"odinger}},\ }\href@noop {} {\bibfield  {journal} {\bibinfo  {journal}
  {Proc. Camb. Phil. Soc.}\ }\textbf {\bibinfo {volume} {31}},\ \bibinfo
  {pages} {553} (\bibinfo {year} {1935})}\BibitemShut {NoStop}%
\bibitem [{\citenamefont {Wiseman}\ \emph {et~al.}(2007)\citenamefont
  {Wiseman}, \citenamefont {Jones},\ and\ \citenamefont {Doherty}}]{wiseman}%
  \BibitemOpen
  \bibfield  {author} {\bibinfo {author} {\bibfnamefont {H.~M.}\ \bibnamefont
  {Wiseman}}, \bibinfo {author} {\bibfnamefont {S.~J.}\ \bibnamefont {Jones}},
  \ and\ \bibinfo {author} {\bibfnamefont {A.~C.}\ \bibnamefont {Doherty}},\
  }\href@noop {} {\bibfield  {journal} {\bibinfo  {journal} {Phys. Rev. Lett.}\
  }\textbf {\bibinfo {volume} {98}},\ \bibinfo {pages} {140402} (\bibinfo
  {year} {2007})}\BibitemShut {NoStop}%
\bibitem [{\citenamefont {Einstein}\ \emph {et~al.}(1935)\citenamefont
  {Einstein}, \citenamefont {Podolsky},\ and\ \citenamefont {Rosen}}]{epr}%
  \BibitemOpen
  \bibfield  {author} {\bibinfo {author} {\bibfnamefont {A.}~\bibnamefont
  {Einstein}}, \bibinfo {author} {\bibfnamefont {B.}~\bibnamefont {Podolsky}},
  \ and\ \bibinfo {author} {\bibfnamefont {N.}~\bibnamefont {Rosen}},\
  }\href@noop {} {\bibfield  {journal} {\bibinfo  {journal} {Phys. Rev.}\
  }\textbf {\bibinfo {volume} {47}},\ \bibinfo {pages} {777} (\bibinfo {year}
  {1935})}\BibitemShut {NoStop}%
\bibitem [{\citenamefont {Reid}(1989)}]{reid}%
  \BibitemOpen
  \bibfield  {author} {\bibinfo {author} {\bibfnamefont {M.~D.}\ \bibnamefont
  {Reid}},\ }\href@noop {} {\bibfield  {journal} {\bibinfo  {journal} {Phys.
  Rev. A}\ }\textbf {\bibinfo {volume} {40}},\ \bibinfo {pages} {913} (\bibinfo
  {year} {1989})}\BibitemShut {NoStop}%
\bibitem [{\citenamefont {Cavalcanti}\ \emph {et~al.}(2009)\citenamefont
  {Cavalcanti}, \citenamefont {Jones}, \citenamefont {Wiseman},\ and\
  \citenamefont {Reid}}]{Cavalcanti_09}%
  \BibitemOpen
  \bibfield  {author} {\bibinfo {author} {\bibfnamefont {E.~G.}\ \bibnamefont
  {Cavalcanti}}, \bibinfo {author} {\bibfnamefont {S.~J.}\ \bibnamefont
  {Jones}}, \bibinfo {author} {\bibfnamefont {H.~M.}\ \bibnamefont {Wiseman}},
  \ and\ \bibinfo {author} {\bibfnamefont {M.~D.}\ \bibnamefont {Reid}},\
  }\href@noop {} {\bibfield  {journal} {\bibinfo  {journal} {Phys. Rev. A.}\
  }\textbf {\bibinfo {volume} {80}},\ \bibinfo {pages} {032112} (\bibinfo
  {year} {2009})}\BibitemShut {NoStop}%
\bibitem [{\citenamefont {Gallego}\ and\ \citenamefont
  {Aolita}(2015)}]{resource}%
  \BibitemOpen
  \bibfield  {author} {\bibinfo {author} {\bibfnamefont {R.}~\bibnamefont
  {Gallego}}\ and\ \bibinfo {author} {\bibfnamefont {L.}~\bibnamefont
  {Aolita}},\ }\href {\doibase10.1103/PhysRevX.5.041008} {\bibfield  {journal}
  {\bibinfo  {journal} {Phys. Rev. X}\ }\textbf {\bibinfo {volume} {5}},\
  \bibinfo {pages} {041008} (\bibinfo {year} {2015})}\BibitemShut {NoStop}%
\bibitem [{\citenamefont {Branciard}\ \emph {et~al.}(2012)\citenamefont
  {Branciard}, \citenamefont {Cavalcanti}, \citenamefont {Walborn},
  \citenamefont {Scarani},\ and\ \citenamefont {Wiseman}}]{branciard}%
  \BibitemOpen
  \bibfield  {author} {\bibinfo {author} {\bibfnamefont {C.}~\bibnamefont
  {Branciard}}, \bibinfo {author} {\bibfnamefont {E.~G.}\ \bibnamefont
  {Cavalcanti}}, \bibinfo {author} {\bibfnamefont {S.~P.}\ \bibnamefont
  {Walborn}}, \bibinfo {author} {\bibfnamefont {V.}~\bibnamefont {Scarani}}, \
  and\ \bibinfo {author} {\bibfnamefont {H.~M.}\ \bibnamefont {Wiseman}},\
  }\href@noop {} {\bibfield  {journal} {\bibinfo  {journal} {Phys. Rev. A}\
  }\textbf {\bibinfo {volume} {85}},\ \bibinfo {pages} {010301} (\bibinfo
  {year} {2012})}\BibitemShut {NoStop}%
\bibitem [{\citenamefont {Walk}\ \emph {et~al.}(2016)\citenamefont {Walk},
  \citenamefont {Hosseini}, \citenamefont {Geng}, \citenamefont {Thearle},
  \citenamefont {Haw}, \citenamefont {Armstrong}, \citenamefont {Assad},
  \citenamefont {Janousek}, \citenamefont {Ralph}, \citenamefont {Symul},
  \citenamefont {Wiseman},\ and\ \citenamefont {Lam}}]{walk}%
  \BibitemOpen
  \bibfield  {author} {\bibinfo {author} {\bibfnamefont {N.}~\bibnamefont
  {Walk}}, \bibinfo {author} {\bibfnamefont {S.}~\bibnamefont {Hosseini}},
  \bibinfo {author} {\bibfnamefont {J.}~\bibnamefont {Geng}}, \bibinfo {author}
  {\bibfnamefont {O.}~\bibnamefont {Thearle}}, \bibinfo {author} {\bibfnamefont
  {J.~Y.}\ \bibnamefont {Haw}}, \bibinfo {author} {\bibfnamefont
  {S.}~\bibnamefont {Armstrong}}, \bibinfo {author} {\bibfnamefont {S.~M.}\
  \bibnamefont {Assad}}, \bibinfo {author} {\bibfnamefont {J.}~\bibnamefont
  {Janousek}}, \bibinfo {author} {\bibfnamefont {T.~C.}\ \bibnamefont {Ralph}},
  \bibinfo {author} {\bibfnamefont {T.}~\bibnamefont {Symul}}, \bibinfo
  {author} {\bibfnamefont {H.~M.}\ \bibnamefont {Wiseman}}, \ and\ \bibinfo
  {author} {\bibfnamefont {P.~K.}\ \bibnamefont {Lam}},\ }\href
  {\doibase10.1364/OPTICA.3.000634} {\bibfield  {journal} {\bibinfo  {journal}
  {Optica}\ }\textbf {\bibinfo {volume} {3}},\ \bibinfo {pages} {634} (\bibinfo
  {year} {2016})}\BibitemShut {NoStop}%
\bibitem [{\citenamefont {Kogias}\ \emph {et~al.}(2017)\citenamefont {Kogias},
  \citenamefont {Xiang}, \citenamefont {He},\ and\ \citenamefont
  {Adesso}}]{kogiasqss}%
  \BibitemOpen
  \bibfield  {author} {\bibinfo {author} {\bibfnamefont {I.}~\bibnamefont
  {Kogias}}, \bibinfo {author} {\bibfnamefont {Y.}~\bibnamefont {Xiang}},
  \bibinfo {author} {\bibfnamefont {Q.~Y.}\ \bibnamefont {He}}, \ and\ \bibinfo
  {author} {\bibfnamefont {G.}~\bibnamefont {Adesso}},\ }\href
  {\doibase10.1103/PhysRevA.95.012315} {\bibfield  {journal} {\bibinfo
  {journal} {Phys. Rev. A}\ }\textbf {\bibinfo {volume} {95}},\ \bibinfo
  {pages} {012315} (\bibinfo {year} {2017})}\BibitemShut {NoStop}%
\bibitem [{\citenamefont {Piani}\ and\ \citenamefont
  {Watrous}(2015)}]{PianiSt}%
  \BibitemOpen
  \bibfield  {author} {\bibinfo {author} {\bibfnamefont {M.}~\bibnamefont
  {Piani}}\ and\ \bibinfo {author} {\bibfnamefont {J.}~\bibnamefont
  {Watrous}},\ }\href {\doibase10.1103/PhysRevLett.114.060404} {\bibfield
  {journal} {\bibinfo  {journal} {Phys. Rev. Lett.}\ }\textbf {\bibinfo
  {volume} {114}},\ \bibinfo {pages} {060404} (\bibinfo {year}
  {2015})}\BibitemShut {NoStop}%
\bibitem [{\citenamefont {Reid}(2013{\natexlab{a}})}]{Reid13}%
  \BibitemOpen
  \bibfield  {author} {\bibinfo {author} {\bibfnamefont {M.~D.}\ \bibnamefont
  {Reid}},\ }\href {\doibase10.1103/PhysRevA.88.062338} {\bibfield  {journal}
  {\bibinfo  {journal} {Phys. Rev. A}\ }\textbf {\bibinfo {volume} {88}},\
  \bibinfo {pages} {062338} (\bibinfo {year} {2013}{\natexlab{a}})}\BibitemShut
  {NoStop}%
\bibitem [{\citenamefont {He}\ \emph {et~al.}(2015{\natexlab{a}})\citenamefont
  {He}, \citenamefont {Rosales-Z\'arate}, \citenamefont {Adesso},\ and\
  \citenamefont {Reid}}]{He15}%
  \BibitemOpen
  \bibfield  {author} {\bibinfo {author} {\bibfnamefont {Q.~Y.}\ \bibnamefont
  {He}}, \bibinfo {author} {\bibfnamefont {L.}~\bibnamefont
  {Rosales-Z\'arate}}, \bibinfo {author} {\bibfnamefont {G.}~\bibnamefont
  {Adesso}}, \ and\ \bibinfo {author} {\bibfnamefont {M.~D.}\ \bibnamefont
  {Reid}},\ }\href {\doibase10.1103/PhysRevLett.115.180502} {\bibfield
  {journal} {\bibinfo  {journal} {Phys. Rev. Lett.}\ }\textbf {\bibinfo
  {volume} {115}},\ \bibinfo {pages} {180502} (\bibinfo {year}
  {2015}{\natexlab{a}})}\BibitemShut {NoStop}%
\bibitem [{\citenamefont {H\"andchen}\ \emph {et~al.}(2012)\citenamefont
  {H\"andchen}, \citenamefont {Eberle}, \citenamefont {Steinlechner},
  \citenamefont {Samblowski}, \citenamefont {Franz}, \citenamefont {Werner},\
  and\ \citenamefont {Schnabel}}]{handchen}%
  \BibitemOpen
  \bibfield  {author} {\bibinfo {author} {\bibfnamefont {V.}~\bibnamefont
  {H\"andchen}}, \bibinfo {author} {\bibfnamefont {T.}~\bibnamefont {Eberle}},
  \bibinfo {author} {\bibfnamefont {S.}~\bibnamefont {Steinlechner}}, \bibinfo
  {author} {\bibfnamefont {A.}~\bibnamefont {Samblowski}}, \bibinfo {author}
  {\bibfnamefont {T.}~\bibnamefont {Franz}}, \bibinfo {author} {\bibfnamefont
  {R.~F.}\ \bibnamefont {Werner}}, \ and\ \bibinfo {author} {\bibfnamefont
  {R.}~\bibnamefont {Schnabel}},\ }\href@noop {} {\bibfield  {journal}
  {\bibinfo  {journal} {Nat. Photon.}\ }\textbf {\bibinfo {volume} {6}},\
  \bibinfo {pages} {596} (\bibinfo {year} {2012})}\BibitemShut {NoStop}%
\bibitem [{\citenamefont {Armstrong}\ \emph {et~al.}(2015)\citenamefont
  {Armstrong}, \citenamefont {Wang}, \citenamefont {Teh}, \citenamefont {Gong},
  \citenamefont {He}, \citenamefont {Janousek}, \citenamefont {Bachor},
  \citenamefont {Reid},\ and\ \citenamefont {Lam}}]{seiji2015}%
  \BibitemOpen
  \bibfield  {author} {\bibinfo {author} {\bibfnamefont {S.}~\bibnamefont
  {Armstrong}}, \bibinfo {author} {\bibfnamefont {M.}~\bibnamefont {Wang}},
  \bibinfo {author} {\bibfnamefont {R.~Y.}\ \bibnamefont {Teh}}, \bibinfo
  {author} {\bibfnamefont {Q.}~\bibnamefont {Gong}}, \bibinfo {author}
  {\bibfnamefont {Q.~Y.}\ \bibnamefont {He}}, \bibinfo {author} {\bibfnamefont
  {J.}~\bibnamefont {Janousek}}, \bibinfo {author} {\bibfnamefont {H.-A.}\
  \bibnamefont {Bachor}}, \bibinfo {author} {\bibfnamefont {M.~D.}\
  \bibnamefont {Reid}}, \ and\ \bibinfo {author} {\bibfnamefont {P.~K.}\
  \bibnamefont {Lam}},\ }\href@noop {} {\bibfield  {journal} {\bibinfo
  {journal} {Nat. Phys.}\ }\textbf {\bibinfo {volume} {11}},\ \bibinfo {pages}
  {167} (\bibinfo {year} {2015})}\BibitemShut {NoStop}%
\bibitem [{\citenamefont {Deng}\ \emph {et~al.}(2017)\citenamefont {Deng},
  \citenamefont {Xiang}, \citenamefont {Tian}, \citenamefont {Adesso},
  \citenamefont {He}, \citenamefont {Gong}, \citenamefont {Su}, \citenamefont
  {Xie},\ and\ \citenamefont {Peng}}]{Deng17}%
  \BibitemOpen
  \bibfield  {author} {\bibinfo {author} {\bibfnamefont {X.}~\bibnamefont
  {Deng}}, \bibinfo {author} {\bibfnamefont {Y.}~\bibnamefont {Xiang}},
  \bibinfo {author} {\bibfnamefont {C.}~\bibnamefont {Tian}}, \bibinfo {author}
  {\bibfnamefont {G.}~\bibnamefont {Adesso}}, \bibinfo {author} {\bibfnamefont
  {Q.~Y.}\ \bibnamefont {He}}, \bibinfo {author} {\bibfnamefont
  {Q.}~\bibnamefont {Gong}}, \bibinfo {author} {\bibfnamefont {X.}~\bibnamefont
  {Su}}, \bibinfo {author} {\bibfnamefont {C.}~\bibnamefont {Xie}}, \ and\
  \bibinfo {author} {\bibfnamefont {K.}~\bibnamefont {Peng}},\ }\href
  {\doibase10.1103/PhysRevLett.118.230501} {\bibfield  {journal} {\bibinfo
  {journal} {Phys. Rev. Lett.}\ }\textbf {\bibinfo {volume} {118}},\ \bibinfo
  {pages} {230501} (\bibinfo {year} {2017})}\BibitemShut {NoStop}%
\bibitem [{\citenamefont {Braunstein}\ and\ \citenamefont {{van
  Loock}}(2005)}]{Brareview}%
  \BibitemOpen
  \bibfield  {author} {\bibinfo {author} {\bibfnamefont {S.~L.}\ \bibnamefont
  {Braunstein}}\ and\ \bibinfo {author} {\bibfnamefont {P.}~\bibnamefont {{van
  Loock}}},\ }\href {\doibase10.1103/RevModPhys.77.513} {\bibfield  {journal}
  {\bibinfo  {journal} {Rev. Mod. Phys.}\ }\textbf {\bibinfo {volume} {77}},\
  \bibinfo {pages} {513} (\bibinfo {year} {2005})}\BibitemShut {NoStop}%
\bibitem [{\citenamefont {Serafini}(2017)}]{serafozzi}%
  \BibitemOpen
  \bibfield  {author} {\bibinfo {author} {\bibfnamefont {A.}~\bibnamefont
  {Serafini}},\ }\href@noop {} {\emph {\bibinfo {title} {Quantum continuous
  variables}}}\ (\bibinfo  {publisher} {Taylor \& Francis, Oxford},\ \bibinfo
  {year} {2017})\BibitemShut {NoStop}%
\bibitem [{\citenamefont {Jones}\ \emph {et~al.}(2007)\citenamefont {Jones},
  \citenamefont {Wiseman},\ and\ \citenamefont {Doherty}}]{wisepra}%
  \BibitemOpen
  \bibfield  {author} {\bibinfo {author} {\bibfnamefont {S.~J.}\ \bibnamefont
  {Jones}}, \bibinfo {author} {\bibfnamefont {H.~M.}\ \bibnamefont {Wiseman}},
  \ and\ \bibinfo {author} {\bibfnamefont {A.~C.}\ \bibnamefont {Doherty}},\
  }\href@noop {} {\bibfield  {journal} {\bibinfo  {journal} {Phys. Rev. A}\
  }\textbf {\bibinfo {volume} {76}},\ \bibinfo {pages} {052116} (\bibinfo
  {year} {2007})}\BibitemShut {NoStop}%
\bibitem [{\citenamefont {He}\ \emph {et~al.}(2015{\natexlab{b}})\citenamefont
  {He}, \citenamefont {Gong},\ and\ \citenamefont {Reid}}]{He2015}%
  \BibitemOpen
  \bibfield  {author} {\bibinfo {author} {\bibfnamefont {Q.~Y.}\ \bibnamefont
  {He}}, \bibinfo {author} {\bibfnamefont {Q.~H.}\ \bibnamefont {Gong}}, \ and\
  \bibinfo {author} {\bibfnamefont {M.~D.}\ \bibnamefont {Reid}},\ }\href
  {\doibase10.1103/PhysRevLett.114.060402} {\bibfield  {journal} {\bibinfo
  {journal} {Phys. Rev. Lett.}\ }\textbf {\bibinfo {volume} {114}},\ \bibinfo
  {pages} {060402} (\bibinfo {year} {2015}{\natexlab{b}})}\BibitemShut
  {NoStop}%
\bibitem [{\citenamefont {Kogias}\ \emph
  {et~al.}(2015{\natexlab{a}})\citenamefont {Kogias}, \citenamefont {Lee},
  \citenamefont {Ragy},\ and\ \citenamefont {Adesso}}]{kogias}%
  \BibitemOpen
  \bibfield  {author} {\bibinfo {author} {\bibfnamefont {I.}~\bibnamefont
  {Kogias}}, \bibinfo {author} {\bibfnamefont {A.}~\bibnamefont {Lee}},
  \bibinfo {author} {\bibfnamefont {S.}~\bibnamefont {Ragy}}, \ and\ \bibinfo
  {author} {\bibfnamefont {G.}~\bibnamefont {Adesso}},\ }\href@noop {}
  {\bibfield  {journal} {\bibinfo  {journal} {Phys. Rev. Lett.}\ }\textbf
  {\bibinfo {volume} {114}},\ \bibinfo {pages} {060403} (\bibinfo {year}
  {2015}{\natexlab{a}})}\BibitemShut {NoStop}%
\bibitem [{\citenamefont {Kogias}\ and\ \citenamefont
  {Adesso}(2015)}]{kogias2}%
  \BibitemOpen
  \bibfield  {author} {\bibinfo {author} {\bibfnamefont {I.}~\bibnamefont
  {Kogias}}\ and\ \bibinfo {author} {\bibfnamefont {G.}~\bibnamefont
  {Adesso}},\ }\href {\doibase10.1364/JOSAB.32.000A27} {\bibfield  {journal}
  {\bibinfo  {journal} {J. Opt. Soc. Am. B}\ }\textbf {\bibinfo {volume}
  {32}},\ \bibinfo {pages} {A27} (\bibinfo {year} {2015})}\BibitemShut
  {NoStop}%
\bibitem [{\citenamefont {Xiang}\ \emph {et~al.}(2017)\citenamefont {Xiang},
  \citenamefont {Kogias}, \citenamefont {Adesso},\ and\ \citenamefont
  {He}}]{kogiasmono}%
  \BibitemOpen
  \bibfield  {author} {\bibinfo {author} {\bibfnamefont {Y.}~\bibnamefont
  {Xiang}}, \bibinfo {author} {\bibfnamefont {I.}~\bibnamefont {Kogias}},
  \bibinfo {author} {\bibfnamefont {G.}~\bibnamefont {Adesso}}, \ and\ \bibinfo
  {author} {\bibfnamefont {Q.~Y.}\ \bibnamefont {He}},\ }\href
  {\doibase10.1103/PhysRevA.95.010101} {\bibfield  {journal} {\bibinfo
  {journal} {Phys. Rev. A}\ }\textbf {\bibinfo {volume} {95}},\ \bibinfo
  {pages} {010101} (\bibinfo {year} {2017})}\BibitemShut {NoStop}%
\bibitem [{\citenamefont {Lami}\ \emph {et~al.}(2016)\citenamefont {Lami},
  \citenamefont {Hirche}, \citenamefont {Adesso},\ and\ \citenamefont
  {Winter}}]{Lami2016}%
  \BibitemOpen
  \bibfield  {author} {\bibinfo {author} {\bibfnamefont {L.}~\bibnamefont
  {Lami}}, \bibinfo {author} {\bibfnamefont {C.}~\bibnamefont {Hirche}},
  \bibinfo {author} {\bibfnamefont {G.}~\bibnamefont {Adesso}}, \ and\ \bibinfo
  {author} {\bibfnamefont {A.}~\bibnamefont {Winter}},\ }\href
  {\doibase10.1103/PhysRevLett.117.220502} {\bibfield  {journal} {\bibinfo
  {journal} {Phys. Rev. Lett.}\ }\textbf {\bibinfo {volume} {117}},\ \bibinfo
  {pages} {220502} (\bibinfo {year} {2016})}\BibitemShut {NoStop}%
\bibitem [{\citenamefont {Adesso}\ and\ \citenamefont
  {Illuminati}(2007)}]{ourreview}%
  \BibitemOpen
  \bibfield  {author} {\bibinfo {author} {\bibfnamefont {G.}~\bibnamefont
  {Adesso}}\ and\ \bibinfo {author} {\bibfnamefont {F.}~\bibnamefont
  {Illuminati}},\ }\href@noop {} {\bibfield  {journal} {\bibinfo  {journal} {J.
  Phys. A: Math. Theor.}\ }\textbf {\bibinfo {volume} {40}},\ \bibinfo {pages}
  {7821} (\bibinfo {year} {2007})}\BibitemShut {NoStop}%
\bibitem [{\citenamefont {Weedbrook}\ \emph {et~al.}(2012)\citenamefont
  {Weedbrook}, \citenamefont {Pirandola}, \citenamefont {Garc\'ia-Patr\'on},
  \citenamefont {Cerf}, \citenamefont {Ralph}, \citenamefont {Shapiro},\ and\
  \citenamefont {Lloyd}}]{pirandolareview}%
  \BibitemOpen
  \bibfield  {author} {\bibinfo {author} {\bibfnamefont {C.}~\bibnamefont
  {Weedbrook}}, \bibinfo {author} {\bibfnamefont {S.}~\bibnamefont
  {Pirandola}}, \bibinfo {author} {\bibfnamefont {R.}~\bibnamefont
  {Garc\'ia-Patr\'on}}, \bibinfo {author} {\bibfnamefont {N.~J.}\ \bibnamefont
  {Cerf}}, \bibinfo {author} {\bibfnamefont {T.~C.}\ \bibnamefont {Ralph}},
  \bibinfo {author} {\bibfnamefont {J.~H.}\ \bibnamefont {Shapiro}}, \ and\
  \bibinfo {author} {\bibfnamefont {S.}~\bibnamefont {Lloyd}},\ }\href@noop {}
  {\bibfield  {journal} {\bibinfo  {journal} {Rev. Mod. Phys.}\ }\textbf
  {\bibinfo {volume} {84}},\ \bibinfo {pages} {621} (\bibinfo {year}
  {2012})}\BibitemShut {NoStop}%
\bibitem [{\citenamefont {Adesso}\ \emph {et~al.}(2014)\citenamefont {Adesso},
  \citenamefont {Ragy},\ and\ \citenamefont {Lee}}]{ournewreview}%
  \BibitemOpen
  \bibfield  {author} {\bibinfo {author} {\bibfnamefont {G.}~\bibnamefont
  {Adesso}}, \bibinfo {author} {\bibfnamefont {S.}~\bibnamefont {Ragy}}, \ and\
  \bibinfo {author} {\bibfnamefont {A.~R.}\ \bibnamefont {Lee}},\ }\href@noop
  {} {\bibfield  {journal} {\bibinfo  {journal} {Open Syst. Inf. Dyn.}\
  }\textbf {\bibinfo {volume} {21}},\ \bibinfo {pages} {1440001} (\bibinfo
  {year} {2014})}\BibitemShut {NoStop}%
\bibitem [{\citenamefont {Menicucci}\ \emph {et~al.}(2006)\citenamefont
  {Menicucci}, \citenamefont {van Loock}, \citenamefont {Gu}, \citenamefont
  {Weedbrook}, \citenamefont {Ralph},\ and\ \citenamefont {Nielsen}}]{CVQC}%
  \BibitemOpen
  \bibfield  {author} {\bibinfo {author} {\bibfnamefont {N.~C.}\ \bibnamefont
  {Menicucci}}, \bibinfo {author} {\bibfnamefont {P.}~\bibnamefont {van
  Loock}}, \bibinfo {author} {\bibfnamefont {M.}~\bibnamefont {Gu}}, \bibinfo
  {author} {\bibfnamefont {C.}~\bibnamefont {Weedbrook}}, \bibinfo {author}
  {\bibfnamefont {T.~C.}\ \bibnamefont {Ralph}}, \ and\ \bibinfo {author}
  {\bibfnamefont {M.~A.}\ \bibnamefont {Nielsen}},\ }\href
  {\doibase10.1103/PhysRevLett.97.110501} {\bibfield  {journal} {\bibinfo
  {journal} {Phys. Rev. Lett.}\ }\textbf {\bibinfo {volume} {97}},\ \bibinfo
  {pages} {110501} (\bibinfo {year} {2006})}\BibitemShut {NoStop}%
\bibitem [{\citenamefont {Wollmann}\ \emph {et~al.}(2016)\citenamefont
  {Wollmann}, \citenamefont {Walk}, \citenamefont {Bennet}, \citenamefont
  {Wiseman},\ and\ \citenamefont {Pryde}}]{OneWayPryde}%
  \BibitemOpen
  \bibfield  {author} {\bibinfo {author} {\bibfnamefont {S.}~\bibnamefont
  {Wollmann}}, \bibinfo {author} {\bibfnamefont {N.}~\bibnamefont {Walk}},
  \bibinfo {author} {\bibfnamefont {A.~J.}\ \bibnamefont {Bennet}}, \bibinfo
  {author} {\bibfnamefont {H.~M.}\ \bibnamefont {Wiseman}}, \ and\ \bibinfo
  {author} {\bibfnamefont {G.~J.}\ \bibnamefont {Pryde}},\ }\href
  {\doibase10.1103/PhysRevLett.116.160403} {\bibfield  {journal} {\bibinfo
  {journal} {Phys. Rev. Lett.}\ }\textbf {\bibinfo {volume} {116}},\ \bibinfo
  {pages} {160403} (\bibinfo {year} {2016})}\BibitemShut {NoStop}%
\bibitem [{\citenamefont {Ji}\ \emph {et~al.}(2016)\citenamefont {Ji},
  \citenamefont {Lee}, \citenamefont {Park},\ and\ \citenamefont
  {Nha}}]{NhaSciRep}%
  \BibitemOpen
  \bibfield  {author} {\bibinfo {author} {\bibfnamefont {S.-W.}\ \bibnamefont
  {Ji}}, \bibinfo {author} {\bibfnamefont {J.}~\bibnamefont {Lee}}, \bibinfo
  {author} {\bibfnamefont {J.}~\bibnamefont {Park}}, \ and\ \bibinfo {author}
  {\bibfnamefont {H.}~\bibnamefont {Nha}},\ }\href {\doibase10.1038/srep29729}
  {\bibfield  {journal} {\bibinfo  {journal} {Sci. Rep.}\ }\textbf {\bibinfo
  {volume} {6}},\ \bibinfo {pages} {29729} (\bibinfo {year}
  {2016})}\BibitemShut {NoStop}%
\bibitem [{\citenamefont {Chen}\ \emph {et~al.}(2002)\citenamefont {Chen},
  \citenamefont {Pan}, \citenamefont {Hou},\ and\ \citenamefont
  {Zhang}}]{chen}%
  \BibitemOpen
  \bibfield  {author} {\bibinfo {author} {\bibfnamefont {Z.~B.}\ \bibnamefont
  {Chen}}, \bibinfo {author} {\bibfnamefont {J.~W.}\ \bibnamefont {Pan}},
  \bibinfo {author} {\bibfnamefont {G.}~\bibnamefont {Hou}}, \ and\ \bibinfo
  {author} {\bibfnamefont {Y.~D.}\ \bibnamefont {Zhang}},\ }\href@noop {}
  {\bibfield  {journal} {\bibinfo  {journal} {Phys. Rev. Lett.}\ }\textbf
  {\bibinfo {volume} {88}},\ \bibinfo {pages} {040406} (\bibinfo {year}
  {2002})}\BibitemShut {NoStop}%
\bibitem [{\citenamefont {Chen}\ and\ \citenamefont {Zhang}(2002)}]{chen2}%
  \BibitemOpen
  \bibfield  {author} {\bibinfo {author} {\bibfnamefont {Z.~B.}\ \bibnamefont
  {Chen}}\ and\ \bibinfo {author} {\bibfnamefont {Y.~D.}\ \bibnamefont
  {Zhang}},\ }\href@noop {} {\bibfield  {journal} {\bibinfo  {journal} {Phys.
  Rev. A}\ }\textbf {\bibinfo {volume} {65}},\ \bibinfo {pages} {044102}
  (\bibinfo {year} {2002})}\BibitemShut {NoStop}%
\bibitem [{\citenamefont {Gour}\ \emph {et~al.}(2004)\citenamefont {Gour},
  \citenamefont {Khanna}, \citenamefont {Mann},\ and\ \citenamefont
  {Revzen}}]{gour1}%
  \BibitemOpen
  \bibfield  {author} {\bibinfo {author} {\bibfnamefont {G.}~\bibnamefont
  {Gour}}, \bibinfo {author} {\bibfnamefont {F.~C.}\ \bibnamefont {Khanna}},
  \bibinfo {author} {\bibfnamefont {A.}~\bibnamefont {Mann}}, \ and\ \bibinfo
  {author} {\bibfnamefont {M.}~\bibnamefont {Revzen}},\ }\href
  {\doibase10.1016/j.physleta.2004.03.018} {\bibfield  {journal} {\bibinfo
  {journal} {Phys. Lett. A}\ }\textbf {\bibinfo {volume} {324}},\ \bibinfo
  {pages} {415} (\bibinfo {year} {2004})}\BibitemShut {NoStop}%
\bibitem [{\citenamefont {Revzen}\ \emph {et~al.}(2005)\citenamefont {Revzen},
  \citenamefont {Mello}, \citenamefont {Mann},\ and\ \citenamefont
  {Johansen}}]{gour2}%
  \BibitemOpen
  \bibfield  {author} {\bibinfo {author} {\bibfnamefont {M.}~\bibnamefont
  {Revzen}}, \bibinfo {author} {\bibfnamefont {P.~A.}\ \bibnamefont {Mello}},
  \bibinfo {author} {\bibfnamefont {A.}~\bibnamefont {Mann}}, \ and\ \bibinfo
  {author} {\bibfnamefont {L.~M.}\ \bibnamefont {Johansen}},\ }\href
  {\doibase10.1103/PhysRevA.71.022103} {\bibfield  {journal} {\bibinfo
  {journal} {Phys. Rev. A}\ }\textbf {\bibinfo {volume} {71}},\ \bibinfo
  {pages} {022103} (\bibinfo {year} {2005})}\BibitemShut {NoStop}%
\bibitem [{\citenamefont {Kogias}\ \emph
  {et~al.}(2015{\natexlab{b}})\citenamefont {Kogias}, \citenamefont
  {Skrzypczyk}, \citenamefont {Cavalcanti}, \citenamefont {Ac\'{\i}n},\ and\
  \citenamefont {Adesso}}]{kogias3}%
  \BibitemOpen
  \bibfield  {author} {\bibinfo {author} {\bibfnamefont {I.}~\bibnamefont
  {Kogias}}, \bibinfo {author} {\bibfnamefont {P.}~\bibnamefont {Skrzypczyk}},
  \bibinfo {author} {\bibfnamefont {D.}~\bibnamefont {Cavalcanti}}, \bibinfo
  {author} {\bibfnamefont {A.}~\bibnamefont {Ac\'{\i}n}}, \ and\ \bibinfo
  {author} {\bibfnamefont {G.}~\bibnamefont {Adesso}},\ }\href
  {\doibase10.1103/PhysRevLett.115.210401} {\bibfield  {journal} {\bibinfo
  {journal} {Phys. Rev. Lett.}\ }\textbf {\bibinfo {volume} {115}},\ \bibinfo
  {pages} {210401} (\bibinfo {year} {2015}{\natexlab{b}})}\BibitemShut
  {NoStop}%
\bibitem [{\citenamefont {{Mi\v{s}ta, Jr.}}\ \emph {et~al.}(2002)\citenamefont
  {{Mi\v{s}ta, Jr.}}, \citenamefont {Filip},\ and\ \citenamefont
  {{Fiur\'{a}\v{s}ek}}}]{Mista_02b}%
  \BibitemOpen
  \bibfield  {author} {\bibinfo {author} {\bibfnamefont {L.}~\bibnamefont
  {{Mi\v{s}ta, Jr.}}}, \bibinfo {author} {\bibfnamefont {R.}~\bibnamefont
  {Filip}}, \ and\ \bibinfo {author} {\bibfnamefont {J.}~\bibnamefont
  {{Fiur\'{a}\v{s}ek}}},\ }\href {\doibase10.1103/PhysRevA.65.062315}
  {\bibfield  {journal} {\bibinfo  {journal} {Phys. Rev. A}\ }\textbf {\bibinfo
  {volume} {65}},\ \bibinfo {pages} {062315} (\bibinfo {year}
  {2002})}\BibitemShut {NoStop}%
\bibitem [{\citenamefont {Simon}\ \emph {et~al.}(1994)\citenamefont {Simon},
  \citenamefont {Mukunda},\ and\ \citenamefont {Dutta}}]{Simon94}%
  \BibitemOpen
  \bibfield  {author} {\bibinfo {author} {\bibfnamefont {R.}~\bibnamefont
  {Simon}}, \bibinfo {author} {\bibfnamefont {N.}~\bibnamefont {Mukunda}}, \
  and\ \bibinfo {author} {\bibfnamefont {B.}~\bibnamefont {Dutta}},\ }\href
  {\doibase10.1103/PhysRevA.49.1567} {\bibfield  {journal} {\bibinfo  {journal}
  {Phys. Rev. A}\ }\textbf {\bibinfo {volume} {49}},\ \bibinfo {pages} {1567}
  (\bibinfo {year} {1994})}\BibitemShut {NoStop}%
\bibitem [{\citenamefont {Simon}(2000)}]{Simon00}%
  \BibitemOpen
  \bibfield  {author} {\bibinfo {author} {\bibfnamefont {R.}~\bibnamefont
  {Simon}},\ }\href@noop {} {\bibfield  {journal} {\bibinfo  {journal} {Phys.
  Rev. Lett.}\ }\textbf {\bibinfo {volume} {84}},\ \bibinfo {pages} {2726}
  (\bibinfo {year} {2000})}\BibitemShut {NoStop}%
\bibitem [{\citenamefont {Duan}\ \emph {et~al.}(2000)\citenamefont {Duan},
  \citenamefont {Giedke}, \citenamefont {Cirac},\ and\ \citenamefont
  {Zoller}}]{Duan00}%
  \BibitemOpen
  \bibfield  {author} {\bibinfo {author} {\bibfnamefont {L.-M.}\ \bibnamefont
  {Duan}}, \bibinfo {author} {\bibfnamefont {G.}~\bibnamefont {Giedke}},
  \bibinfo {author} {\bibfnamefont {J.~I.}\ \bibnamefont {Cirac}}, \ and\
  \bibinfo {author} {\bibfnamefont {P.}~\bibnamefont {Zoller}},\ }\href
  {\doibase10.1103/PhysRevLett.84.2722} {\bibfield  {journal} {\bibinfo
  {journal} {Phys. Rev. Lett.}\ }\textbf {\bibinfo {volume} {84}},\ \bibinfo
  {pages} {2722} (\bibinfo {year} {2000})}\BibitemShut {NoStop}%
\bibitem [{\citenamefont {Adesso}\ \emph {et~al.}(2004)\citenamefont {Adesso},
  \citenamefont {Serafini},\ and\ \citenamefont {Illuminati}}]{extremal}%
  \BibitemOpen
  \bibfield  {author} {\bibinfo {author} {\bibfnamefont {G.}~\bibnamefont
  {Adesso}}, \bibinfo {author} {\bibfnamefont {A.}~\bibnamefont {Serafini}}, \
  and\ \bibinfo {author} {\bibfnamefont {F.}~\bibnamefont {Illuminati}},\
  }\href@noop {} {\bibfield  {journal} {\bibinfo  {journal} {Phys. Rev. A}\
  }\textbf {\bibinfo {volume} {70}},\ \bibinfo {pages} {022318} (\bibinfo
  {year} {2004})}\BibitemShut {NoStop}%
\bibitem [{\citenamefont {Filip}\ and\ \citenamefont {{Mi\v{s}ta,
  Jr.}}(2002)}]{Mista_02}%
  \BibitemOpen
  \bibfield  {author} {\bibinfo {author} {\bibfnamefont {R.}~\bibnamefont
  {Filip}}\ and\ \bibinfo {author} {\bibfnamefont {L.}~\bibnamefont
  {{Mi\v{s}ta, Jr.}}},\ }\href {\doibase10.1103/PhysRevA.66.044309} {\bibfield
  {journal} {\bibinfo  {journal} {Phys. Rev. A}\ }\textbf {\bibinfo {volume}
  {66}},\ \bibinfo {pages} {044309} (\bibinfo {year} {2002})}\BibitemShut
  {NoStop}%
\bibitem [{\citenamefont {Xu}\ \emph {et~al.}(2017)\citenamefont {Xu},
  \citenamefont {Tufarelli},\ and\ \citenamefont {Adesso}}]{Buqing}%
  \BibitemOpen
  \bibfield  {author} {\bibinfo {author} {\bibfnamefont {B.}~\bibnamefont
  {Xu}}, \bibinfo {author} {\bibfnamefont {T.}~\bibnamefont {Tufarelli}}, \
  and\ \bibinfo {author} {\bibfnamefont {G.}~\bibnamefont {Adesso}},\ }\href
  {\doibase10.1103/PhysRevA.95.012124} {\bibfield  {journal} {\bibinfo
  {journal} {Phys. Rev. A}\ }\textbf {\bibinfo {volume} {95}},\ \bibinfo
  {pages} {012124} (\bibinfo {year} {2017})}\BibitemShut {NoStop}%
\bibitem [{\citenamefont {Ferraro}\ and\ \citenamefont
  {Paris}(2005)}]{ferraro23}%
  \BibitemOpen
  \bibfield  {author} {\bibinfo {author} {\bibfnamefont {A.}~\bibnamefont
  {Ferraro}}\ and\ \bibinfo {author} {\bibfnamefont {M.~G.~A.}\ \bibnamefont
  {Paris}},\ }\href {\doibase10.1088/1464-4266/7/6/003} {\bibfield  {journal}
  {\bibinfo  {journal} {Journal of Optics B: Quantum and Semiclassical Optics}\
  }\textbf {\bibinfo {volume} {7}},\ \bibinfo {pages} {174} (\bibinfo {year}
  {2005})}\BibitemShut {NoStop}%
\bibitem [{\citenamefont {Martin}\ and\ \citenamefont {Vennin}(2017)}]{vennin}%
  \BibitemOpen
  \bibfield  {author} {\bibinfo {author} {\bibfnamefont {J.}~\bibnamefont
  {Martin}}\ and\ \bibinfo {author} {\bibfnamefont {V.}~\bibnamefont
  {Vennin}},\ }\href@noop {} {\bibfield  {journal} {\bibinfo  {journal}
  {arXiv:1706.05001}\ } (\bibinfo {year} {2017})}\BibitemShut {NoStop}%
\bibitem [{\citenamefont {Pirandola}\ \emph {et~al.}(2014)\citenamefont
  {Pirandola}, \citenamefont {Spedalieri}, \citenamefont {Braunstein},
  \citenamefont {Cerf},\ and\ \citenamefont {Lloyd}}]{Pirandola_14}%
  \BibitemOpen
  \bibfield  {author} {\bibinfo {author} {\bibfnamefont {S.}~\bibnamefont
  {Pirandola}}, \bibinfo {author} {\bibfnamefont {G.}~\bibnamefont
  {Spedalieri}}, \bibinfo {author} {\bibfnamefont {S.~L.}\ \bibnamefont
  {Braunstein}}, \bibinfo {author} {\bibfnamefont {N.~J.}\ \bibnamefont
  {Cerf}}, \ and\ \bibinfo {author} {\bibfnamefont {S.}~\bibnamefont {Lloyd}},\
  }\href {\doibase10.1103/PhysRevLett.113.140405} {\bibfield  {journal}
  {\bibinfo  {journal} {Phys. Rev. Lett.}\ }\textbf {\bibinfo {volume} {113}},\
  \bibinfo {pages} {140405} (\bibinfo {year} {2014})}\BibitemShut {NoStop}%
\bibitem [{\citenamefont {Caruso}\ \emph {et~al.}(2006)\citenamefont {Caruso},
  \citenamefont {Giovannetti},\ and\ \citenamefont {Holevo}}]{Caruso_06}%
  \BibitemOpen
  \bibfield  {author} {\bibinfo {author} {\bibfnamefont {F.}~\bibnamefont
  {Caruso}}, \bibinfo {author} {\bibfnamefont {V.}~\bibnamefont {Giovannetti}},
  \ and\ \bibinfo {author} {\bibfnamefont {A.~S.}\ \bibnamefont {Holevo}},\
  }\href@noop {} {\bibfield  {journal} {\bibinfo  {journal} {New J. Phys.}\
  }\textbf {\bibinfo {volume} {8}},\ \bibinfo {pages} {310} (\bibinfo {year}
  {2006})}\BibitemShut {NoStop}%
\bibitem [{\citenamefont {Garc\'{\i}a-Patr\'{o}n}\ \emph
  {et~al.}(2012)\citenamefont {Garc\'{\i}a-Patr\'{o}n}, \citenamefont
  {Navarrete-Benlloch}, \citenamefont {Lloyd}, \citenamefont {Shapiro},\ and\
  \citenamefont {Cerf}}]{Garcia-Patron_12}%
  \BibitemOpen
  \bibfield  {author} {\bibinfo {author} {\bibfnamefont {R.}~\bibnamefont
  {Garc\'{\i}a-Patr\'{o}n}}, \bibinfo {author} {\bibfnamefont {C.}~\bibnamefont
  {Navarrete-Benlloch}}, \bibinfo {author} {\bibfnamefont {S.}~\bibnamefont
  {Lloyd}}, \bibinfo {author} {\bibfnamefont {J.~H.}\ \bibnamefont {Shapiro}},
  \ and\ \bibinfo {author} {\bibfnamefont {N.~J.}\ \bibnamefont {Cerf}},\
  }\href@noop {} {\bibfield  {journal} {\bibinfo  {journal} {Phys. Rev. Lett.}\
  }\textbf {\bibinfo {volume} {108}},\ \bibinfo {pages} {110505} (\bibinfo
  {year} {2012})}\BibitemShut {NoStop}%
\bibitem [{\citenamefont {{Mi\v{s}ta, Jr.}}\ \emph {et~al.}(2014)\citenamefont
  {{Mi\v{s}ta, Jr.}}, \citenamefont {McNulty},\ and\ \citenamefont
  {Adesso}}]{Mista_14}%
  \BibitemOpen
  \bibfield  {author} {\bibinfo {author} {\bibfnamefont {L.}~\bibnamefont
  {{Mi\v{s}ta, Jr.}}}, \bibinfo {author} {\bibfnamefont {D.}~\bibnamefont
  {McNulty}}, \ and\ \bibinfo {author} {\bibfnamefont {G.}~\bibnamefont
  {Adesso}},\ }\href {\doibase10.1103/PhysRevA.90.022328} {\bibfield  {journal}
  {\bibinfo  {journal} {Phys. Rev. A}\ }\textbf {\bibinfo {volume} {90}},\
  \bibinfo {pages} {022328} (\bibinfo {year} {2014})}\BibitemShut {NoStop}%
\bibitem [{\citenamefont {Kraus}\ and\ \citenamefont {Cirac}(2004)}]{Kraus_04}%
  \BibitemOpen
  \bibfield  {author} {\bibinfo {author} {\bibfnamefont {B.}~\bibnamefont
  {Kraus}}\ and\ \bibinfo {author} {\bibfnamefont {J.~I.}\ \bibnamefont
  {Cirac}},\ }\href {\doibase10.1103/PhysRevLett.92.013602} {\bibfield
  {journal} {\bibinfo  {journal} {Phys. Rev. Lett.}\ }\textbf {\bibinfo
  {volume} {92}},\ \bibinfo {pages} {013602} (\bibinfo {year}
  {2004})}\BibitemShut {NoStop}%
\bibitem [{\citenamefont {Paternostro}\ \emph {et~al.}(2004)\citenamefont
  {Paternostro}, \citenamefont {Son},\ and\ \citenamefont
  {Kim}}]{Paternostro_04}%
  \BibitemOpen
  \bibfield  {author} {\bibinfo {author} {\bibfnamefont {M.}~\bibnamefont
  {Paternostro}}, \bibinfo {author} {\bibfnamefont {W.}~\bibnamefont {Son}}, \
  and\ \bibinfo {author} {\bibfnamefont {M.~S.}\ \bibnamefont {Kim}},\ }\href
  {\doibase10.1103/PhysRevLett.92.197901} {\bibfield  {journal} {\bibinfo
  {journal} {Phys. Rev. Lett.}\ }\textbf {\bibinfo {volume} {92}},\ \bibinfo
  {pages} {197901} (\bibinfo {year} {2004})}\BibitemShut {NoStop}%
\bibitem [{\citenamefont {Adesso}\ \emph {et~al.}(2010)\citenamefont {Adesso},
  \citenamefont {Campbell}, \citenamefont {Illuminati},\ and\ \citenamefont
  {Paternostro}}]{Paternostro_10}%
  \BibitemOpen
  \bibfield  {author} {\bibinfo {author} {\bibfnamefont {G.}~\bibnamefont
  {Adesso}}, \bibinfo {author} {\bibfnamefont {S.}~\bibnamefont {Campbell}},
  \bibinfo {author} {\bibfnamefont {F.}~\bibnamefont {Illuminati}}, \ and\
  \bibinfo {author} {\bibfnamefont {M.}~\bibnamefont {Paternostro}},\ }\href
  {\doibase10.1103/PhysRevLett.104.240501} {\bibfield  {journal} {\bibinfo
  {journal} {Phys. Rev. Lett.}\ }\textbf {\bibinfo {volume} {104}},\ \bibinfo
  {pages} {240501} (\bibinfo {year} {2010})}\BibitemShut {NoStop}%
\bibitem [{\citenamefont {Paternostro}\ \emph {et~al.}(2009)\citenamefont
  {Paternostro}, \citenamefont {Adesso},\ and\ \citenamefont
  {Campbell}}]{Passing_09}%
  \BibitemOpen
  \bibfield  {author} {\bibinfo {author} {\bibfnamefont {M.}~\bibnamefont
  {Paternostro}}, \bibinfo {author} {\bibfnamefont {G.}~\bibnamefont {Adesso}},
  \ and\ \bibinfo {author} {\bibfnamefont {S.}~\bibnamefont {Campbell}},\
  }\href {\doibase10.1103/PhysRevA.80.062318} {\bibfield  {journal} {\bibinfo
  {journal} {Phys. Rev. A}\ }\textbf {\bibinfo {volume} {80}},\ \bibinfo
  {pages} {062318} (\bibinfo {year} {2009})}\BibitemShut {NoStop}%
\bibitem [{\citenamefont {Jones}\ and\ \citenamefont
  {Wiseman}(2011)}]{HowardCri}%
  \BibitemOpen
  \bibfield  {author} {\bibinfo {author} {\bibfnamefont {S.~J.}\ \bibnamefont
  {Jones}}\ and\ \bibinfo {author} {\bibfnamefont {H.~M.}\ \bibnamefont
  {Wiseman}},\ }\href@noop {} {\bibfield  {journal} {\bibinfo  {journal} {Phys.
  Rev. A.}\ }\textbf {\bibinfo {volume} {84}},\ \bibinfo {pages} {012110}
  (\bibinfo {year} {2011})}\BibitemShut {NoStop}%
\bibitem [{\citenamefont {Tatham}\ \emph {et~al.}(2012)\citenamefont {Tatham},
  \citenamefont {{Mi\v{s}ta, Jr.}}, \citenamefont {Adesso},\ and\ \citenamefont
  {Korolkova}}]{Tatham_12}%
  \BibitemOpen
  \bibfield  {author} {\bibinfo {author} {\bibfnamefont {R.}~\bibnamefont
  {Tatham}}, \bibinfo {author} {\bibfnamefont {L.}~\bibnamefont {{Mi\v{s}ta,
  Jr.}}}, \bibinfo {author} {\bibfnamefont {G.}~\bibnamefont {Adesso}}, \ and\
  \bibinfo {author} {\bibfnamefont {N.}~\bibnamefont {Korolkova}},\ }\href
  {\doibase10.1103/PhysRevA.85.022326} {\bibfield  {journal} {\bibinfo
  {journal} {Phys. Rev. A}\ }\textbf {\bibinfo {volume} {85}},\ \bibinfo
  {pages} {022326} (\bibinfo {year} {2012})}\BibitemShut {NoStop}%
\bibitem [{\citenamefont {Werner}(1989)}]{Werner_89}%
  \BibitemOpen
  \bibfield  {author} {\bibinfo {author} {\bibfnamefont {R.~F.}\ \bibnamefont
  {Werner}},\ }\href {\doibase10.1103/PhysRevA.40.4277} {\bibfield  {journal}
  {\bibinfo  {journal} {Phys. Rev. A}\ }\textbf {\bibinfo {volume} {40}},\
  \bibinfo {pages} {4277} (\bibinfo {year} {1989})}\BibitemShut {NoStop}%
\bibitem [{\citenamefont {He}\ and\ \citenamefont {Reid}(2013)}]{he}%
  \BibitemOpen
  \bibfield  {author} {\bibinfo {author} {\bibfnamefont {Q.~Y.}\ \bibnamefont
  {He}}\ and\ \bibinfo {author} {\bibfnamefont {M.~D.}\ \bibnamefont {Reid}},\
  }\href@noop {} {\bibfield  {journal} {\bibinfo  {journal} {Phys. Rev. Lett.}\
  }\textbf {\bibinfo {volume} {111}},\ \bibinfo {pages} {250403} (\bibinfo
  {year} {2013})}\BibitemShut {NoStop}%
\bibitem [{\citenamefont {Cavalcanti}\ \emph {et~al.}(2015)\citenamefont
  {Cavalcanti}, \citenamefont {Skrzypczyk}, \citenamefont {Aguilar},
  \citenamefont {Nery}, \citenamefont {Ribeiro},\ and\ \citenamefont
  {Walborn}}]{cavalcanti-nc}%
  \BibitemOpen
  \bibfield  {author} {\bibinfo {author} {\bibfnamefont {D.}~\bibnamefont
  {Cavalcanti}}, \bibinfo {author} {\bibfnamefont {P.}~\bibnamefont
  {Skrzypczyk}}, \bibinfo {author} {\bibfnamefont {G.}~\bibnamefont {Aguilar}},
  \bibinfo {author} {\bibfnamefont {R.}~\bibnamefont {Nery}}, \bibinfo {author}
  {\bibfnamefont {P.~S.}\ \bibnamefont {Ribeiro}}, \ and\ \bibinfo {author}
  {\bibfnamefont {S.}~\bibnamefont {Walborn}},\ }\href
  {http://dx.doi.org/10.1038/ncomms8941} {\bibfield  {journal} {\bibinfo
  {journal} {Nat. Commun.}\ }\textbf {\bibinfo {volume} {6}},\ \bibinfo {pages}
  {7941} (\bibinfo {year} {2015})}\BibitemShut {NoStop}%
\bibitem [{\citenamefont {Reid}(2013{\natexlab{b}})}]{reid13a}%
  \BibitemOpen
  \bibfield  {author} {\bibinfo {author} {\bibfnamefont {M.~D.}\ \bibnamefont
  {Reid}},\ }\href@noop {} {\bibfield  {journal} {\bibinfo  {journal} {Phys.
  Rev. A}\ }\textbf {\bibinfo {volume} {88}},\ \bibinfo {pages} {062108}
  (\bibinfo {year} {2013}{\natexlab{b}})}\BibitemShut {NoStop}%
\bibitem [{\citenamefont {Dodonov}\ \emph {et~al.}(1984)\citenamefont
  {Dodonov}, \citenamefont {Man'ko},\ and\ \citenamefont
  {Semjonov}}]{Dodonov_84}%
  \BibitemOpen
  \bibfield  {author} {\bibinfo {author} {\bibfnamefont {V.~V.}\ \bibnamefont
  {Dodonov}}, \bibinfo {author} {\bibfnamefont {V.~I.}\ \bibnamefont {Man'ko}},
  \ and\ \bibinfo {author} {\bibfnamefont {V.~V.}\ \bibnamefont {Semjonov}},\
  }\href@noop {} {\bibfield  {journal} {\bibinfo  {journal} {Nuovo Cimento B}\
  }\textbf {\bibinfo {volume} {83}},\ \bibinfo {pages} {145} (\bibinfo {year}
  {1984})}\BibitemShut {NoStop}%
\bibitem [{\citenamefont {Dodonov}\ \emph {et~al.}(1994)\citenamefont
  {Dodonov}, \citenamefont {Man'ko},\ and\ \citenamefont
  {Man'ko}}]{Dodonov_94}%
  \BibitemOpen
  \bibfield  {author} {\bibinfo {author} {\bibfnamefont {V.~V.}\ \bibnamefont
  {Dodonov}}, \bibinfo {author} {\bibfnamefont {O.~V.}\ \bibnamefont {Man'ko}},
  \ and\ \bibinfo {author} {\bibfnamefont {V.~I.}\ \bibnamefont {Man'ko}},\
  }\href@noop {} {\bibfield  {journal} {\bibinfo  {journal} {Phys. Rev. A}\
  }\textbf {\bibinfo {volume} {50}},\ \bibinfo {pages} {813} (\bibinfo {year}
  {1994})}\BibitemShut {NoStop}%
\bibitem [{\citenamefont {Fiur\'{a}\v{s}ek}\ and\ \citenamefont
  {Pe\v{r}ina}(2001)}]{Fiurasek_review01}%
  \BibitemOpen
  \bibfield  {author} {\bibinfo {author} {\bibfnamefont {J.}~\bibnamefont
  {Fiur\'{a}\v{s}ek}}\ and\ \bibinfo {author} {\bibfnamefont {J.}~\bibnamefont
  {Pe\v{r}ina}},\ }in\ \href@noop {} {\emph {\bibinfo {booktitle} {Coherence
  and Statistics of Photons and Atoms}}},\ \bibinfo {editor} {edited by\
  \bibinfo {editor} {\bibfnamefont {J.}~\bibnamefont {Pe\v{r}ina}}}\ (\bibinfo
  {publisher} {J. Wiley},\ \bibinfo {address} {New York},\ \bibinfo {year}
  {2001})\ Chap.~\bibinfo {chapter} {2}, pp.\ \bibinfo {pages}
  {65--110}\BibitemShut {NoStop}%
\bibitem [{\citenamefont {Erd\'elyi}(1953)}]{Bateman_53}%
  \BibitemOpen
  \bibinfo {editor} {\bibfnamefont {A.}~\bibnamefont {Erd\'elyi}},\ ed.,\
  \href@noop {} {\emph {\bibinfo {title} {{\it Bateman Manuscript Project:}
  {\it Higher Transcendental Functions}}}}\ (\bibinfo  {publisher}
  {McGraw-Hill},\ \bibinfo {address} {New York},\ \bibinfo {year}
  {1953})\BibitemShut {NoStop}%
\end{thebibliography}

\end{document}